\documentclass[twocolumn]{aastex701}

\newcommand{\dphi}{\Delta \phi}
\newcommand{\aamp}{5.6 \times 10^{-3}}
\newcommand{\aamperr}{1.5 \times 10^{-3}}
\newcommand{\apow}{5.4 \times 10^{-1}}
\newcommand{\apowerr}{4.6 \times 10^{-2}}

\newcommand{\bamp}{1.3 \times 10^{-3}}
\newcommand{\bamperr}{3.3 \times 10^{-4}}
\newcommand{\bpow}{5.4 \times 10^{-1}}
\newcommand{\bpowerr}{4.6 \times 10^{-2}}

\begin{document}

%%\title{Template \aastex v7.0.1 Article with Examples\footnote{Footnotes can be added to titles}}

\title[Planetary Systems Along Stellar Streams]{Life is But a Stream: The  Distribution of Planetary Systems Along Stellar Streams and their Properties}

\author[sname='Webb']{Jeremy J. Webb}
\affiliation{Department of Science, Technology, and Society, Division of Natural Sciences, York University, 218 Bethune College, Toronto, ON, M3J1P3}
\affiliation{David A. Dunlap Department of Astronomy \& Astrophysics, University of Toronto 50 St. George St, Toronto M5S 3H4, Canada}
\email[show]{webbjj@yorku.ca}  

\author[gname=Milica, sname='Ivetic']{Milica Ivetic} 
\affiliation{David A. Dunlap Department of Astronomy \& Astrophysics, University of Toronto 50 St. George St, Toronto M5S 3H4, Canada}
\email[show]{milica.ivetic@alumni.utoronto.ca}

\author[gname=Maxwell,sname=Cai]{Maxwell X. Cai}
\affiliation{Leiden Observatory, Leiden University, P.O. Box 9513, 2300 RA, Leiden, The Netherlands}
\affiliation{Cavendish Laboratory, University of Cambridge, J.J. Thomson Ave, Cambridge CB3 OH3, United Kingdom}
\email[show]{mxc22@cam.ac.uk}

\author{Simon Portegies Zwart}
\affiliation{Leiden Observatory, Leiden University, P.O. Box 9513, 2300 RA, Leiden, The Netherlands}
\email{spz@strw.leidenuniv.nl}

\author{Daniella Morrone} 
\affiliation{David A. Dunlap Department of Astronomy \& Astrophysics, University of Toronto 50 St. George St, Toronto M5S 3H4, Canada}
\email{daniella.morrone@utoronto.ca}

\begin{abstract}

The majority of discovered exoplanets have been observed orbiting field stars as opposed to within a star cluster. To determine whether the lack of observed exoplanets in star clusters is due to gravitational perturbations or observational limitations, we consider the possibility of studying exoplanets in stellar streams. We present the results of direct $N$-body simulations of planetary systems around stars that orbit within a star cluster. Our simulations demonstrate that stars with early cluster escape times tend to retain all their planets as they spend most of their time orbiting in the cluster's low-density outskirts. Alternatively, stars with later escape times can have a wide range of survival fractions as they are subjected to a range of local densities and encounter types. With respect to the stellar stream that forms as the result of the cluster's dissolution, stars near the edge of the stream are therefore more likely to have unperturbed planetary systems. Conversely, stars near the centre of the stream have a higher chance of having planets pushed to eccentric orbits, inclined orbits, or stripped from the system entirely. From our suite of simulations, we provide an estimate of the probability that a star will host a planet with a given initial semi-major axis $a_0$ based on the star's location along a stellar stream $\Delta \phi$.

\end{abstract}



\section{Introduction}

The collapse of a molecular cloud leads to stars forming in clustered environments that can host between several tens to several millions of stars \citep{Lada2003,PortegiesZwart2010}. Lower mass systems, commonly referred to as OB Associations, will break apart quickly. Higher mass systems, or star clusters, dissolve over much longer timescales \citep{PortegiesZwart2010}. In either case, the fact that stars do not form alone means that any planetary systems that form around stars will also spend time in a clustered environment.

Planets have been found to be quite prevalent in the Universe, with observational surveys finding main sequence stars with masses between 0.5-1.2 $M_{\odot}$ are likely to host at least one planet \citep{Cassan2012, Winn2015}. However, the majority of planets orbiting stars other than our Sun (exoplanets) have been observed in the field as opposed to within a star cluster. Less than 100 planets have been detected in star clusters \citep{Cai2019, Dai2023}, with only one planet discovered in a dense globular cluster \citep{Backer1993,Sigurdsson1993}.

While a planetary system's host star remains in its birth association or cluster, planets may be subjected to gravitational perturbations due to passing stars. Such perturbations can alter the orbital distribution of individual planets or remove them entirely from the system \citep{vanElteren2019,Cai2019,DaffernPowell2022}. Planets that are perturbed to the point of escaping their host star would become free-floating planets within the birth cluster, comparable to the hundreds of free floating planets observed throughout the Orion Nebula \citep{Drass2016} and other star forming regions \citep{Luhman07, Scholz2012, Martin2025}. Hence one possible explanation for the lack of exoplanet system detections in star clusters is that planetary systems do not survive long in dense clustered environments. This solution would suggest that the majority of field exoplanets were either born in loosely bound OB associations that broke apart quickly or escaped their host cluster not long after formation, and only a small fraction spend extensive amounts of time within a star cluster. An alternative solution would be that it is difficult to detect exoplanet systems within star clusters due to crowding effects, with the stellar density being too high to isolate individual stars.

The properties of the Solar System, and the fact that life exists on Earth, have been used to place constraints on the Sun's birth cluster. The pollution of primitive meteorites  by nearby supernova \citet{PortegiesZwart2021, Arzoumanian2023}, the tilt of the ecliptic with respect to the Sun's equatorial plane, and the structure of the Kuiper belt can be used to constrain how close other stars have passed to our Sun. \citet{PortegiesZwart2009b} used $N$-body simulations to estimate the Sun's birth cluster to have been within 500-3000 $M_{\odot}$ with a radius of 1-3 pc in order to reproduce these observations. Additional constraints, like limiting background radiation exposure to preserve life on Earth and the dynamical scattering of Sedna \citep{Hanse2018} by a close encounter, are also in agreement with this prediction \citep{Adams2010}. Several studies have used these properties to search for the lost solar siblings of the Sun to better understand its birth environment, however matching field stars to the same progenitor birth cluster requires both chemical and kinematic information for all stars and contains a significant degree of uncertainty associated with Galactic potential \citep[e.g][]{Brown2010, Batista2012, Batista2014, Adibekyan2018, Webb2020, Pfalzner2020}.

Studying exoplanets in stellar streams offers a unique test of whether or not the lack of observed exoplanets in star clusters is due to gravitational perturbations or observational limitations. Throughout a cluster's evolution, internal (stellar evolution, two-body relaxation) and external (tidal stripping, shocks) mechanisms eventually lead the cluster to dissolve into a stellar stream \citep{Spitzer1958, Spitzer1987, Heggie2003}. A stellar stream's density will be considerably lower than its progenitor cluster, hence it will be easier to detect exoplanets in stellar streams if they have formed and remain bound to their host star due to minimal crowding effects from nearby stars. In fact, \citet{Newton2021} discovered a three-planet system orbiting around a young star in the Pisces-Eridanus stream. Furthermore, streams have been shown to retain dynamical knowledge of their progenitor cluster. A star's location along the stream is indicative of when it escaped its host cluster, with stars that escape early being located near the edges of the stream while the stream's centre is populated by the last stars to escape the cluster. Mass segregation along a stream reveals the progenitor cluster's relaxation rate, with the stellar mass function at a star's location along a stream reflecting the host cluster's dynamical state at the time of escape \citep{Ye2021, Pang2021, Webb2022}. 

Given how a star's location along a stream is linked to its escape time and its progenitor's dynamical history, it is implied that a star's location along a stream will also reflect the stellar environment that it experienced while bound to its host cluster. Stars that escape their host cluster early likely have spent most of their time in the cluster's outskirts, where the stellar density is low. Conversely, star's that escape the cluster late will likely have spent some time in the dense stellar core. \citet{Cai2019} found that a star's ability to retain a planet with a semi-major axis $a$ depends strongly on the mean stellar density that the host star experiences. Therefore, when looking along a stellar stream, we would expect stars near the edges of a stream to have in-tact and minimally perturbed planetary systems and stars near the centre of stream to have heavily perturbed planetary systems.

In this study, we consider how the properties of planetary systems that evolve within a star cluster are distributed along the resulting stellar stream that remains after the host cluster has dissolved. Given a direct $N$-body simulation of a star cluster, we further simulate the evolution of identical planetary systems around all Sun-like stars in the cluster until the star becomes energetically unbound. An analysis of how the properties of each planetary system has changed as a function of the star's location along the stream will provide a prediction for the probability of finding an exoplanet around a given stream star. The analysis will further provide insight into whether or not observed exoplanets could only have come from loosely bound stellar associations. In Section \ref{sec:method} we explain how we simulated the evolution of a large number of planetary systems orbiting stars embedded within a star cluster. The results of our simulations are presented in Section \ref{sec:results}, while the probability of planetary system survival and how much its initial state has been altered are discussed in Section \ref{sec:discussion}. Finally, we summarize our findings in Section \ref{sec:conclusion}.

\section{Method} \label{sec:method}

To simulate the planetary systems around selected host stars we use \textsc{LonelyPlanets} \citep{2017MNRAS.470.4337C, 2018MNRAS.474.5114C, Cai2019}, which utilizes the the \textsc{AMUSE} (Astrophysical Multipurpose Software Environment) framework \footnote{https://www.amusecode.org/} \cite{PortegiesZwart2009, PortegiesZwart2013, Pelupessy2013}. AMUSE allows for separate numerical integrators to pass relevant information between them. The framework, first introduced in \citet{2017MNRAS.470.4337C}, allows for a star cluster-scale simulation to be done separately from a planetary system-scale simulation while still allowing for information to pass between the two. The first integrator that we use is the direct N-body code \textsc{NBODY6++GPU} code\citep{Aarseth2003, Wang2015}, which is commonly used for modelling star cluster evolution. A NBODY6++GPU star cluster simulation allows us to simulate the local cluster environment around individual planet-hosting stars based on high-frequency output of stellar positions and velocities (1000 year time steps) \citep{2015ApJS..219...31C}. \textsc{LonelyPlanets} then determines the relative positions and velocities of a given host star's five nearest neighbours at any time between the NBODY6++GPU timesteps. 

The second integrator that we use is IAS15 \citep{Rein2015} within REBOUND \citep{Rein2012}, to simulate the evolution of planets around a host star. Planetary systems are evolved separately due to the relevant spatial scales and timescales of planetary systems being several orders of magnitude smaller than those of a star cluster. In addition to the massless planets orbiting around a host star, the relative positions and velocities of nearby cluster stars are passed from the first integrator to REBOUND, allowing for these neighbors to gravitationally perturb planets. We simulate the evolution of each planetary systems until the cluster fully dissolved.

The methodology behind our approach and the details of both our star cluster and planetary system simulations are outlined in the following subsections.

\subsection{Star Cluster Simulation}

We simulate a cluster of 12500 stars with an initial mass of 7500 $M_{\odot}$ and an initial half-mass radius of 4pc. %\textbf{This initial cluster is more massive and longer than the canonical birth cluster given in REF, but is suitable at our adopted distance to the galactic center of 5kpc.} \todo[inline]{JW - Can't find reference, plus suitable how? As in the dissolution time?} 
The stellar initial mass function is taken from \citet{Kroupa2001} and the initial density profile is a \citet{Plummer1911} sphere. The cluster is given a circular orbit at 5 kpc in a Milky Way-like tidal field (MWPOTENTIAL2014 from \cite{Bovy2015}). The system evolved until its complete dissolution, which we denote to be the time when the cluster has less than 100 energetically bound stars remaining. Given this criteria, cluster dissolution occurs after 1088 Myr, with a long thin stream remaining.

%\todo[inline]{SPZ: This is more massive and longer than the Kanomical Beith cluster according toportegies zwart 2020, but suitable at our adopted distance to the galactic center at 5kpc}

\begin{figure*}
    \includegraphics[width=\textwidth]{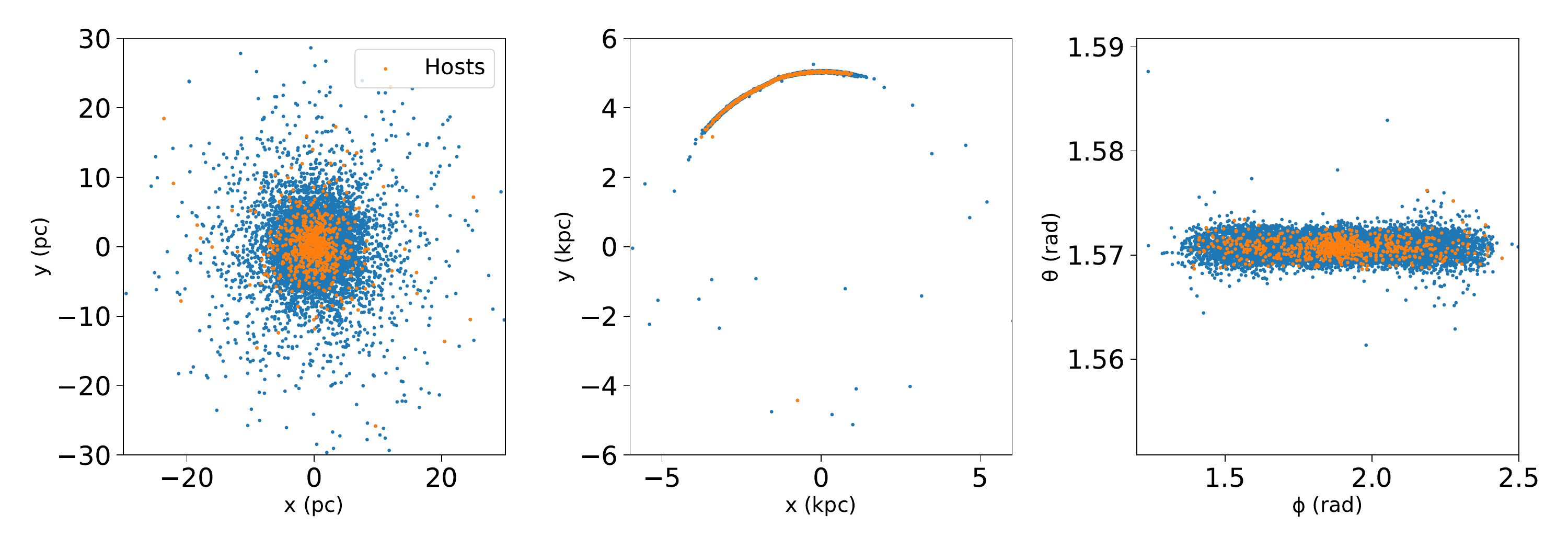}
    \caption{Initial projected positions of cluster stars in clustercentric coordinates (left panel), final projected positions of cluster stars in Galactocentric cartesian coordinates after the cluster has dissolved into a stellar stream (middle panel), and angular positions of cluster stars in Galactocentric spherical coordinates after the cluster has dissolved into a stellar stream (right panel). In each panel, the orange points denote stars around which we simulate the evolution of a planetary system.}
    \label{fig:gcplot}
\end{figure*}

To study how planetary systems evolve around Sun-like stars in this cluster, we select all main sequence stars with masses between 0.8-1.2 $M_{\odot}$ to be host stars. Figure \ref{fig:gcplot} illustrates the initial and final positions of stars in the cluster, with the 705 Sun-like host stars marked in orange.

\begin{figure}
    \includegraphics[width=0.48\textwidth]{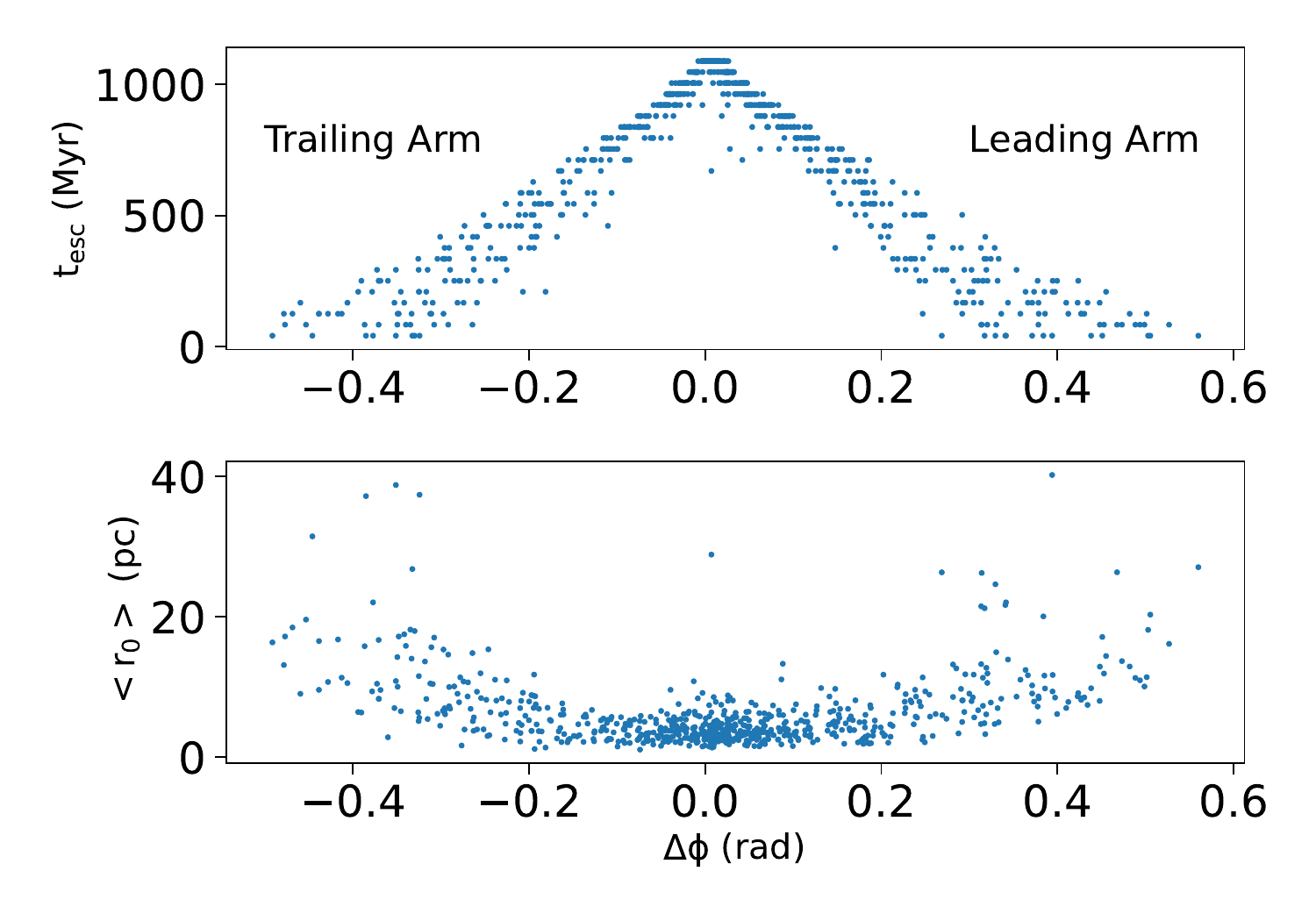}
    \caption{Escape time (top panel) and initial mean orbital distance (bottom panel) of host stars as a function of the angular separations between each star and the centre of the stream. A star's $\dphi$ is an excellent proxy for when it escaped the host cluster and loosely correlated with each star's initial orbit within the host cluster. }
    \label{fig:phi_tdiss}
\end{figure}

Once the cluster has dissolved, we calculate the angular separation between each host star and the centre of the stream $\dphi$, with the centre being the location that the progenitor would have had it not dissolved. The upper panel of Figure \ref{fig:phi_tdiss} expectedly demonstrates that $\dphi$ is closely linked to a star's escape time, with the first stars to escape the cluster being located near the edges of the stream where the local stellar density falls to the background density. Conversely, the last stars to escape the cluster are located near the centre. For each host star, we also calculate the initial mean orbital distance $\langle r_0 \rangle$, which is calculated by finding the time-averaged orbital distance of each star during the cluster's first two cluster dynamical timescales. The lower panel of Figure \ref{fig:phi_tdiss} further demonstrates that $\dphi$ is also an indicator of $\langle r_0 \rangle$. Hence a star's location along the stream is a proxy for the stellar environment the star experienced while bound to the cluster and the duration of time the star spent in this environment. However, it should be noted that as $\dphi$ approaches zero there is a range of $\langle r_0 \rangle$ values that stars can have.

It is interesting to note in Figure \ref{fig:phi_tdiss} the existence of one star with an intermediate escape time (544 Myr) and a large $r_0$ ($\sim$ 30 pc) despite being located near the centre of the stream ($\dphi = 0.01$). This star became energetically unbound from the cluster after 544 Myr and was never recaptured, but spent its lifetime on an orbit that keeps it close to the progenitor cluster. The possible existence of unbound stars that remain close to their host cluster for long periods of time was pointed out by \citep{Henon1970}. It is also worth noting that in denser clusters, stars that experience strong encounters in the cluster's core can get ejected from their host cluster and contaminate its tidal tails \citep{Grondin2024, Weatherford2026}. Hence there will be outliers where a star's location along the stream is not truly indicative of its escape time or initial orbit in the host cluster, but given our simulation these cases appear to be rare.

Considering a suite of cluster orbits, masses, sizes, and density profiles all serve to change the range of escape times, mean orbital distances, and densities experienced by individual stars. The relationships of these parameters with location along the stream will also differ. While this study is applicable to typical open cluster streams \citep[e.g.]{Curtis2019, Ye2021} and provides insight into the Sun's birth cluster, additional simulations are required to extend this analysis to the tidal tails of older open clusters and globular clusters \citep{Piatti2020, Bonaca2025}. However, the wide range of local densities and encounter scenarios experienced by stars in our model will also occur in clusters with different initial properties.

\subsection{Planetary System Simulations}

To study how planetary systems evolve around Sun-like stars in the cluster, we first initialize 14 massless planets with initial eccentricities and inclinations of zero around each host star. Since the planets are massless, they can only be perturbed by nearby cluster stars and not via planet-planet scattering. The initial semi-major axes of the two innermost planets is set to 10 au and 20 au, followed by 9 planets evenly spaced out at 20 au intervals. To probe the stability of the outer regions of each star's planetary system, we also consider planets with orbits at 600 au, 1000 au, and 2000 au. These semi-major axis values allow us to test the stability of a wide range of orbits while minimizing the computational expense of the simulations.

\section{Results}\label{sec:results}

In order to quantify how a planetary system has been perturbed from its initial state due to close encounters with other clusters, we calculate the survival fraction $f$ of each planetary system, the final orbital eccentricity $e$ of surviving planets, and the ratio of current to initial in orbital energy $E/E_0$ of surviving planets.

\subsection{Survival}

%\todo[inline]{MXC: Personally I found that Fig 3 & 4 are difficult to understand, because they basically suggest that f is irrelevant to $\dphi$. Indeed, the survivability also depends on $t_{esc}$. How about using $t_{esc}$ as the y-axis and color code the survival rates instead? Also, it is unexplained why the number of dots is the smallest with $f = 0.5$.}
%\todo[inline]{JW: To me the purpose of these plots is to show what to expect for an observer, who isn't going to know $\dphi$. So we establish in Figure \ref{fig:phi_tdiss} that $\dphi$ is a decent proxy for $t_{esc}$ and then go from there. Maybe that needs to be explicitly stated?}

\begin{figure}
    \includegraphics[width=0.48\textwidth]{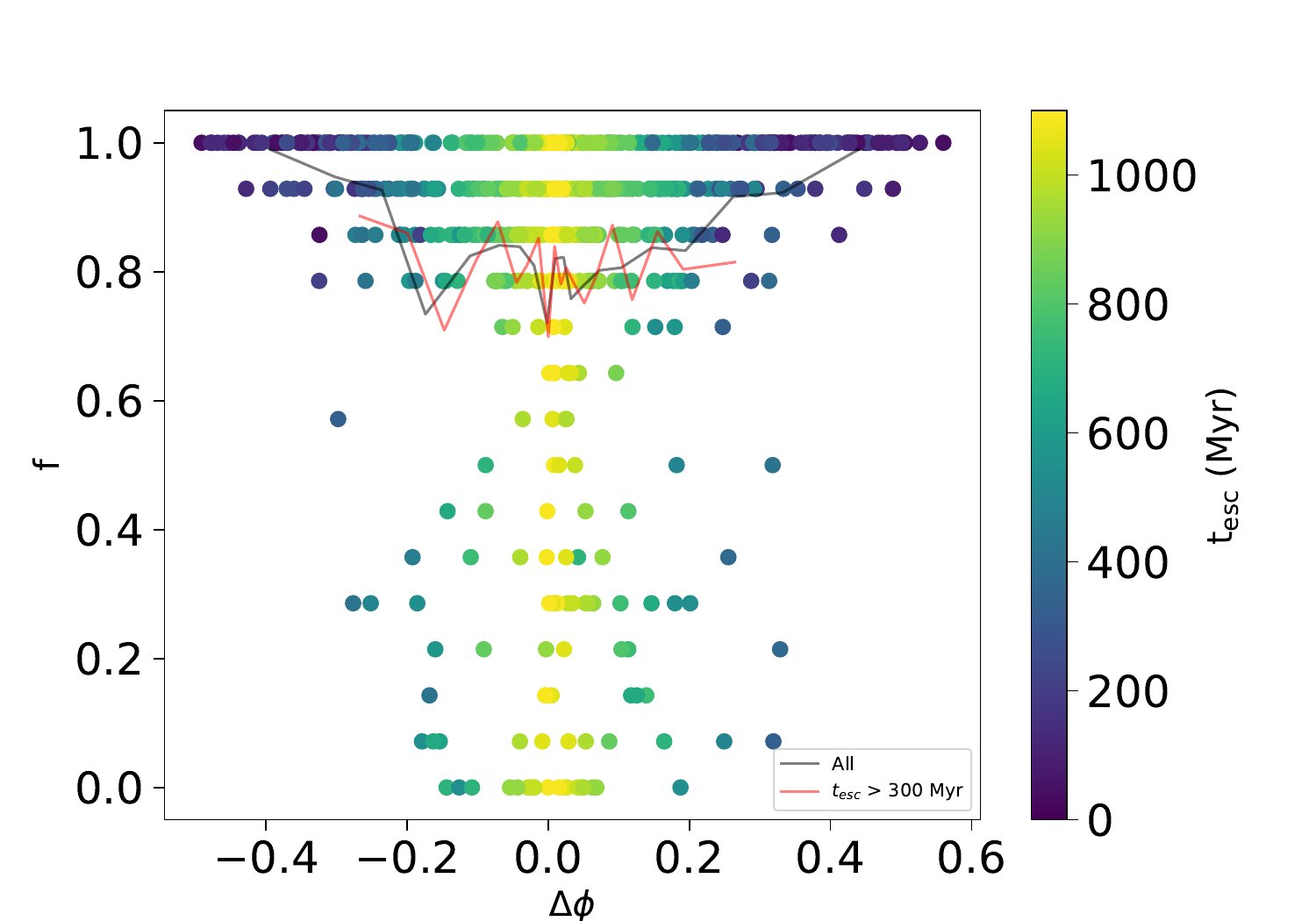}
    \caption{Fraction of planets that remain bound to their host star as a function of the host star's angular separation $\dphi$ from the stream's centre after cluster dissolution. Points are colour-coded by each stars escape time for the cluster. Solid lines represent the mean survival fraction as a function of $\dphi$ for all stars (grey) and stars that escape after 300 Myr (red). Host stars that escape the cluster early, and are located at the edges of the stream, have a greater chance of retaining most or all of their planets. Host stars that escape the cluster late, and are located near the centre of the stream, have a range of survival fractions as stars are more likely to undergo close encounters that strip planets from the system the longer they remain within the cluster.}
    \label{fig:survival_phi}
\end{figure}

Figure \ref{fig:survival_phi} illustrates how after the cluster dissolves the survival fraction $f$ of each planetary system is related to the host star's location along the stellar stream. Points in Figure \ref{fig:survival_phi} are colour coded by the host star's escape time, which in Figure \ref{fig:phi_tdiss} we demonstrated is related to the star's angular offset from the stream centre $\dphi$. Host stars at the edges of the stream, with larger $|\dphi|$ values, all appear to have survival fractions near 1. This result is consistent with Figure \ref{fig:phi_tdiss} in that these stars escape the cluster early and have a larger mean orbital distance. More specifically, stars with $|\dphi| > 0.2$ escape the cluster within the first 300 Myr of evolution and have initial mean orbital distances greater than 8pc. Hence these stars have a low-probability of having a close encounter with a nearby star given the low local density and the short amount of time they spend in the cluster. It is interesting to note that 300 Myr corresponds to twice the cluster's orbital period around the Galaxy and the cluster's half-mass relaxation time.

For $|\dphi| < 0.2$, the stream is populated by stars that spend more time in the cluster and therefore experience both a wide range of density environments and have a higher chance of a close encounter. These stars can have a range of survival fractions, with some systems retaining the majority of their planets while others are stripped completely. Separating the systems into those with high survival fractions ($f > 0.8$) and low survival fractions ($f < 0.8$), we find the dispersion in the $\dphi$ of systems with high survival fractions (0.22) is twice that of systems with low survival fractions (0.11).

\begin{figure}
    \includegraphics[width=0.48\textwidth]{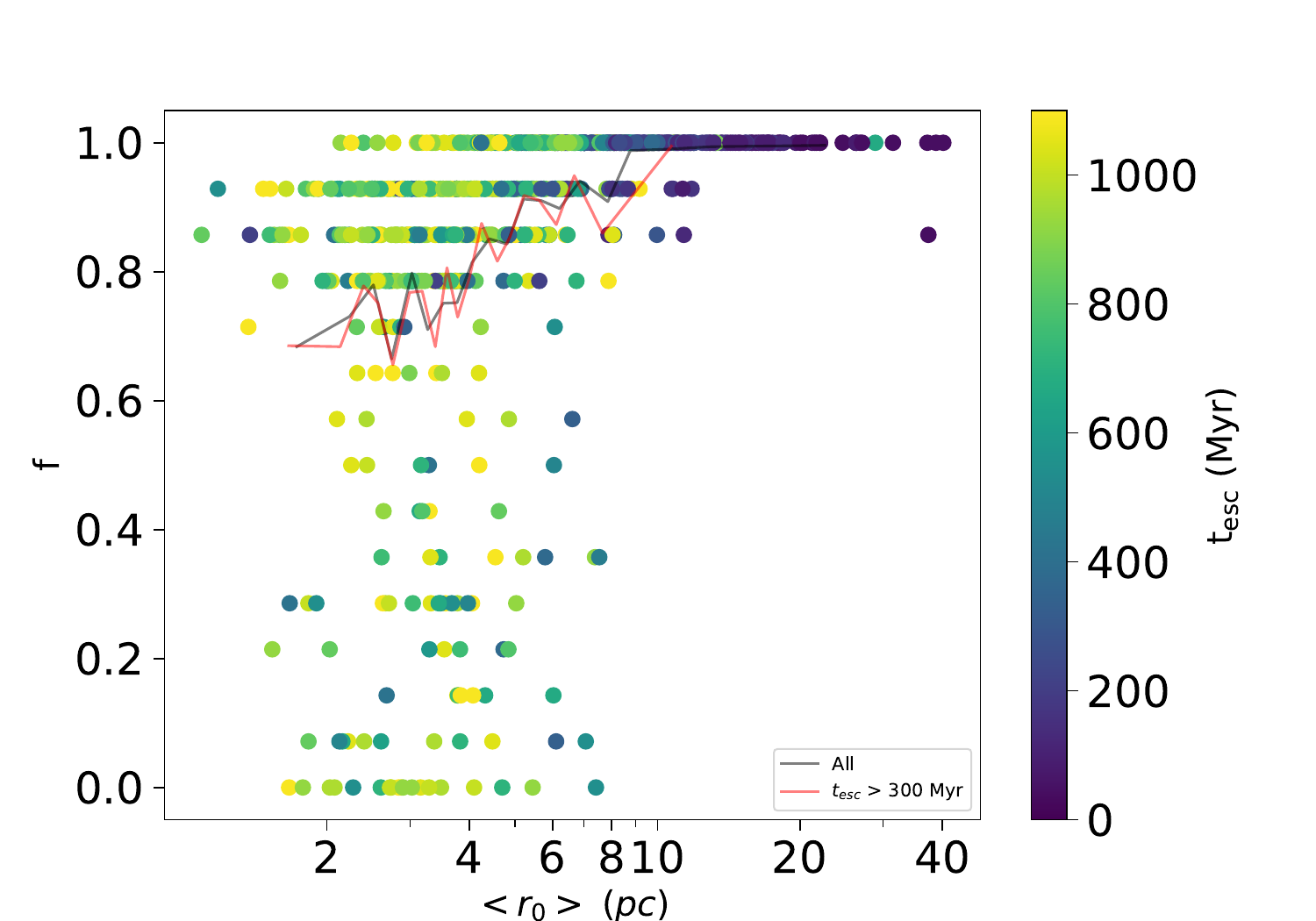}
    \caption{Fraction of planets that remain bound to their host star as a function of the host star's initial mean orbital distance. Points are colour-coded by each stars escape time for the cluster. Solid lines represent the mean survival fraction as a function of $<r_0>$ for all stars (grey) and stars that escape after 300 Myr (red). Host stars that initially orbit in the outskirts of the cluster have a greater chance of retaining most or all of their planets. Host stars that initially orbit in the inner regions of the cluster have a range of survival fractions as stars are more likely to undergo close encounters that strip planets from the system when orbiting in denser environments.}
    \label{fig:survival_r0}
\end{figure}

The wide range of possible $f$ values for stars with $|\dphi| < 0.2$ can be partially understood by looking at the initial orbits of these stars in the cluster's potential in Figure \ref{fig:survival_r0}. Stars that escape the cluster late are not necessarily also stars that were initially found in the core of the cluster. Many appear to have initial mean orbital distances beyond the cluster's half-mass radius. Hence it is possible for a star to spend the majority of its lifetime in a low density environment despite remaining bound to the cluster until the cluster reaches dissolution. Similarly, some stars can have initial orbits in the inner regions of the cluster and retain most of their planets despite having late escape times. Therefore, while an early escape time seems to nearly guaranty a host star to retain all its planets, a late escape time can lead to a wide range of dynamical histories and hence a wide range of survival fractions.

%\todo[inline]{JW: I don't think anything more quantitative needs to be said here given the analysis in the discussion, but am open to suggestions. We did do a KS-test to determine if the $\dphi$ distribution is different for systems with f>0.8 and systems with f<0.8. They are! Worth including?}
%\todo[inline]{SPZ: Would be good to quantify these statements though}
%\todo[inline]{JW: What about dispersion's in phi2 for f>0.8 and f<0.8, as above.}

\subsection{Orbital Perturbation}

\begin{figure*}
    \includegraphics[width=1\textwidth]{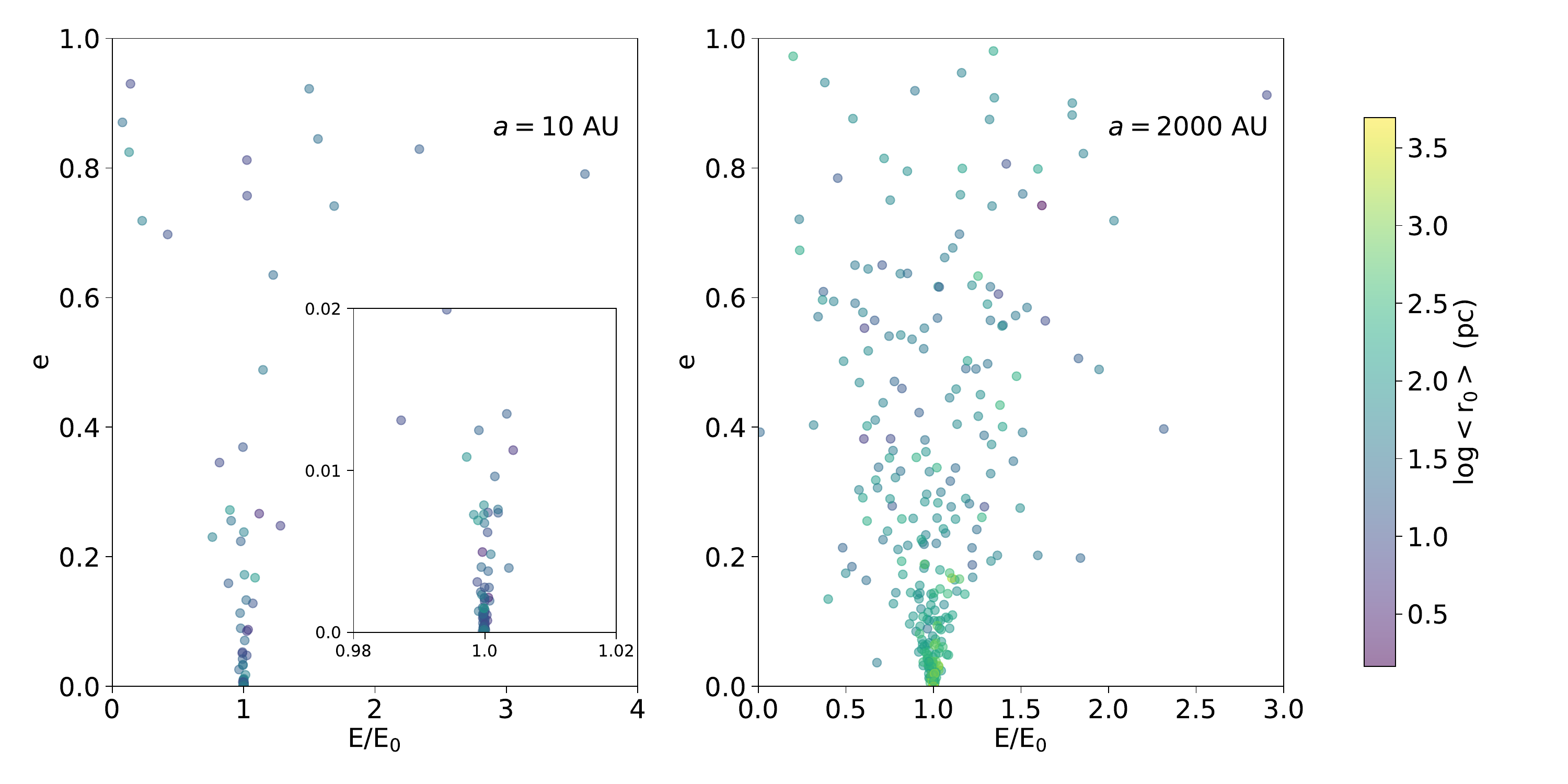}
    \caption{The change in orbital eccentricity $e-e_0$ compared to the ratio of current to initial in orbital energy $E/E_0$ of surviving planets with an initial semi-major axis of 10 au (left panel) and 2000 au (right panel). Points are colour-coded by the natural logarithm of the host star's initial mean orbital distance. The left panel inset focuses on the region near each planet's initial location of $e=0$ and $E/E_0=1$. The most common case for inner planets is to remain close to their initial orbit, even if their host star orbits in a relatively dense environment. However there are several cases, regardless of initial mean orbital distance, where stars have likely undergone a close encounter that perturb the inner most planets in our simulations to new orbits. The most common case for outer planets is to be perturbed from their original orbit, with the degree of perturbation scaling with initial mean orbital distance.}
    \label{fig:ecc_energya2}
\end{figure*}

Having demonstrated how orbiting within a star cluster can affect the survival fraction of a host star's initial planetary system, we next consider the state of the star's surviving planets. More specifically, we explore whether the surviving planets have been perturbed from their initial orbit or not. For the innermost and outermost planets in our simulations we consider their final eccentricity $e$ and the ratio of current to initial in orbital energy $E/E_0$ in Figure \ref{fig:ecc_energya2}. In the figure, points are colour coded by the nautral logarithm of the host star's initial mean orbital distance within the star cluster.

The left panel of Figure \ref{fig:ecc_energya2} illustrates that for planets with circular orbits at an initial semi-major axis of 10 au, their final orbital eccentricity and energy are quite close to their initial values, since nearly $90\%$ of the surviving planets are located within the zoomed in inset. However for $10\%$ of the cases, the planets orbit has changed by a non-negligible amount due to one or more perturbations by passing cluster stars. The degree of perturbation does not strongly depend on the host stars initial mean orbital distance, suggesting that for inner planets that have been strongly perturbed it only takes a few strong perturbations to significantly alter the orbit of the innermost planets in our simulations. It is also more likely that strong perturbations will happen early in a cluster's lifetime when the cluster is most dense, suggesting that after some critical time in a cluster's evolution, inner planets will be safe from being perturbed from their initial orbit.

In the right panel of Figure \ref{fig:ecc_energya2} we see that planets with circular orbits at an initial semi-major axis of 2000 au are more easily perturbed by the cluster environment. For these planets, only $10\%$ of the cases would be within the parameter space outlined by the left panel's inset. Most planets have been significantly perturbed from their initial orbital energy and eccentricity. Converse to the 10 au case, it appears that host stars with a small initial mean orbital distance have their outermost planets more significantly perturbed than for systems with large initial mean orbital distances.

\section{Discussion}\label{sec:discussion}

The star cluster simulation and suite of planetary system simulations presented in this study provide insight into whether or not planetary systems can survive in star clusters as they dissolve into a stellar stream. The simulations indicate that planetary systems are able to survive in clustered environments for an extended period of time, with the probability of survival being strongly dependent on the host star's initial orbit in the cluster and loosely dependent on the host star's escape time. For planetary systems that form in star clusters comparable to our model cluster, the simulations suggest that many are expected to survive and populate the resulting stellar stream. This result would suggest that the lack of detected exoplanets in star clusters is due to the difficulty of finding exoplanets around stars in high-density environments. It is instead easier to find exoplanets in the lower-density end products of star cluster evolution, stellar streams, which we find provide a potential window into the evolution of planetary systems in star clusters. In the following subsection, we quantify the probability of finding an exoplanet with a given semi-major axis around stellar stream stars.

\subsection{Probability of Survival}
\begin{figure*}
    \includegraphics[width=\textwidth]{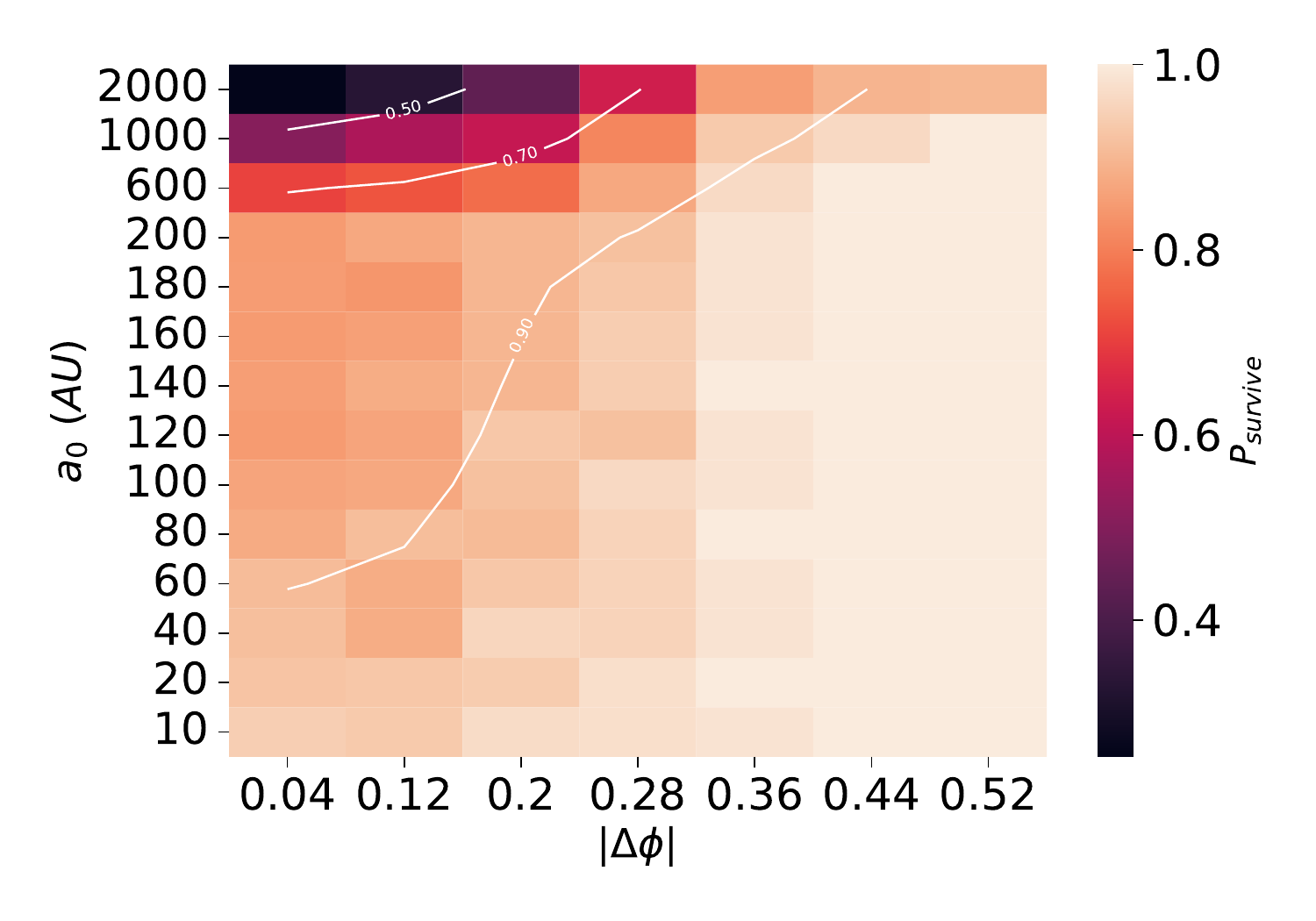}
    \caption{Probability of a planet with initial semi major axis $a_0$ remaining bound to its host star after the birth cluster dissolves as a function of absolute value of the angular separation $|\Delta \phi|$ from the centre of the stream. Contours of $P_{\rm survive}$ are determined after applying a multidimensional Gaussian filter to the dataset with a standard deviation of 1.  Stars close to the centre of the stream are less likely to host planets at any $a_0$, with outer planets more likely to be stripped from their host star than inner planet. Stars near the edges of the stream are more likely to retain planets out to 2000 au.}
    \label{fig:dphi_survival}
\end{figure*}

To estimate the probability that a star with a given $\dphi$ will host a planet that formed with in initial semi-major axis $a_0$, we first bin the host stars based on their respective $\dphi$. For all host stars in a given bin, we find the fraction of planets with a given $a_0$ that are still bound to their host star at the end of the simulation. This fraction is then set equal to the probability that a planet will remain bound to its host star after the birth cluster dissolves. The results of these calculations are illustrated in Figure \ref{fig:dphi_survival}.

Figure \ref{fig:dphi_survival} reveals how strongly the probability of survival depends on $a_0$ and $\dphi$. Near the centre of the stream, a 2000 au planet only remains bound to its host cluster $20\%$ of the time, with the probability of survival increasing as $\dphi$ increases. At the edges of the stream, in the outermost $\dphi$ bin, all planets within 1000 au remain bound to their host stars. For a fixed $\dphi$, the probability that a planet remains bound to its host star decreases as $a_0$ increases. These results suggest that the probability of a planet with a given $a_0$ remaining bound to its host star after the birth cluster has dissolved can be predicted as a function of $\dphi$. We consider the probability of survival $P_{\rm survive}$ having a power-law dependence on $\dphi$ in the form of:

\begin{equation}\label{eqn:prob}
\log_{10}(P_{\rm survive})=\alpha(a_0) \log_{10}(|\Delta \phi|) + \beta(a_0).
\end{equation}

\begin{figure*}
    \includegraphics[width=\textwidth]{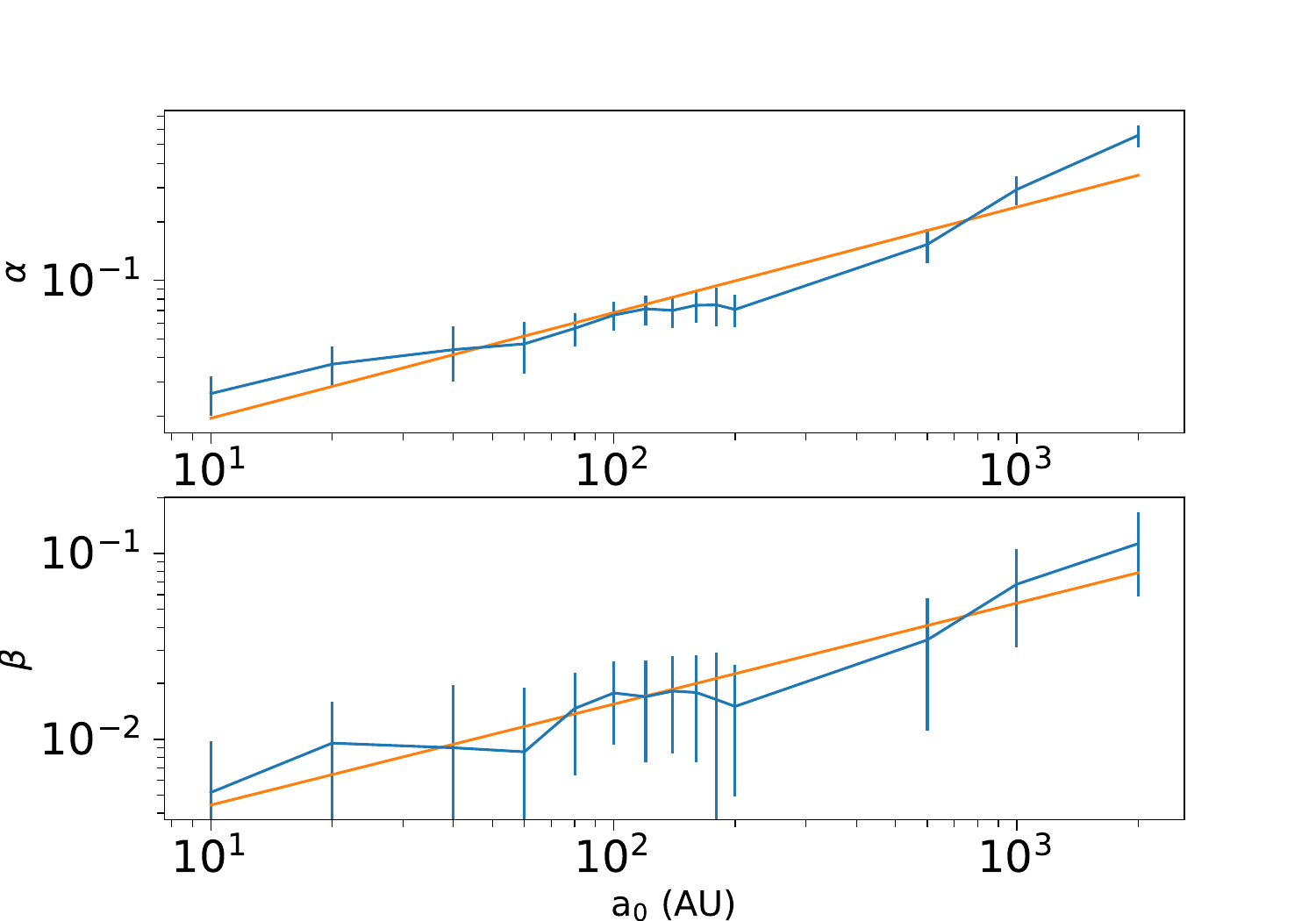}
    \caption{Dependence of $\alpha$ (top panel) and $\beta$ (bottom panel) in Equation \ref{eqn:prob} on a planet's initial semi-major axis $a_0$. A power-law is fit to each relationship, with the best fit parameters given in Equations \ref{eqn:alpha} and \ref{eqn:beta}.}
    \label{fig:alpha_beta}
\end{figure*}

Fitting Equation \ref{eqn:prob} for separate values of $a_0$ reveals that $\alpha$ and $\beta$ are each power-law functions of $a_0$, as illustrated in Figure \ref{fig:alpha_beta}. The best fitting power-law functions to the two relationships are:

\begin{equation}\label{eqn:alpha}
\alpha = (\aamp \pm \aamperr) \ a_0 ^ {\apow \pm \apowerr},
\end{equation}

\begin{equation}\label{eqn:beta}
\beta = (\bamp \pm \bamperr) \ a_0 ^ {\bpow \pm \bpowerr}.
\end{equation}

More simply, we can combine Equations \ref{eqn:prob}-\ref{eqn:beta} to be:

\begin{equation}
    P_{\rm survive}=A^{a_0^B} (\Delta \phi)^{C a_0^D}
\end{equation}

\noindent with $A=1.003 \pm 0.008$, $B=\bpow \pm \bpowerr$, $C=\aamp \pm \aamperr$, and $D=\apow \pm \apowerr$.

\begin{figure}
    \includegraphics[width=0.48\textwidth]{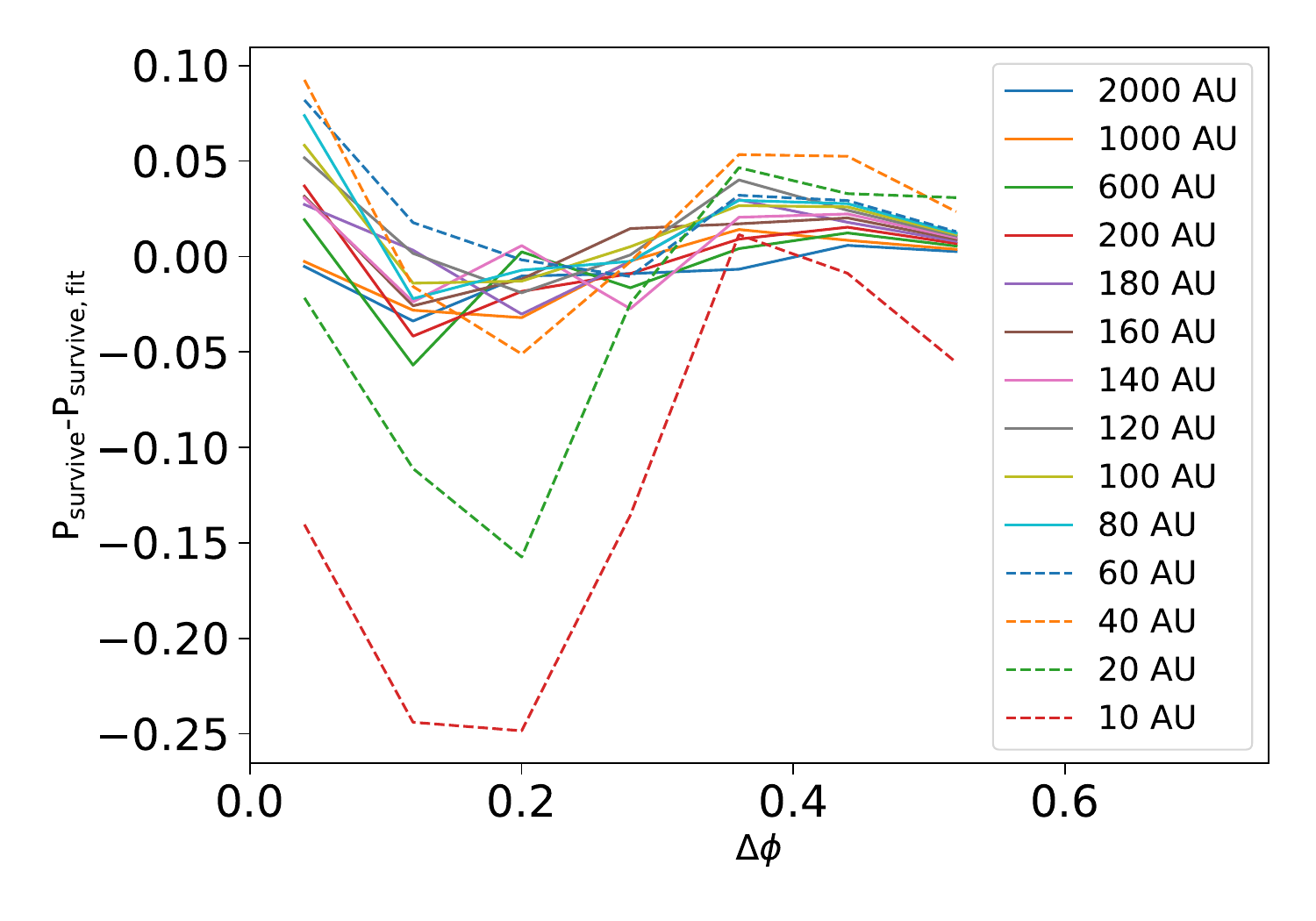}
    \caption{Residuals between measured probability of survival $P_{\rm survive}$ and fitted probability of survival $P_{\rm survive,fit}$ as per Equation \ref{eqn:prob}.}
    \label{fig:fit_res}
\end{figure}

To understand the accuracy in which Equations \ref{eqn:prob}-\ref{eqn:beta} estimate the probability that a planet with semi-major axis $a_0$ will remain bound to a host star at a location $|\Delta \phi|$ from the stream's centre, we consider the residuals between the measured probability of survival $P_{\rm survive}$ in Figure \ref{fig:dphi_survival} and the fitted probability of survival $P_{\rm survive,fit}$. This comparison, illustrated in Figure \ref{fig:fit_res}, reveals that Equations \ref{eqn:prob}-\ref{eqn:beta} can predict the probability of survival for planets with initial semi-axis greater than 20 au to within 10 $\%$. The accuracy is much higher for outer region planets, where the power-law fits to $\alpha$ and $\beta$ match the simulated data. For lower values of $a_0$ the Equations \ref{eqn:prob}-\ref{eqn:beta} have difficulty predicting $P_{\rm survive}$. However for $a_0$ values less than or equal to 40 au, the probability of survival is greater than 80 $\%$ no matter where the planet is located along the stream. Hence, for the model cluster considered here, planets within 40 au of their host star have a high probability of remaining bound.

It is important to note that Equations \ref{eqn:prob}-\ref{eqn:beta} have been determined using a single model cluster. Given the strong dependence of planet survival on the mean stellar density it experiences \citep{Cai2019}, extending this work to include denser clusters and clusters with longer dissolution times will be necessary to determine how the Equations apply to old open cluster and globular cluster streams.

\subsection{Degree of Perturbation}

\begin{figure}
    \includegraphics[width=0.48\textwidth]{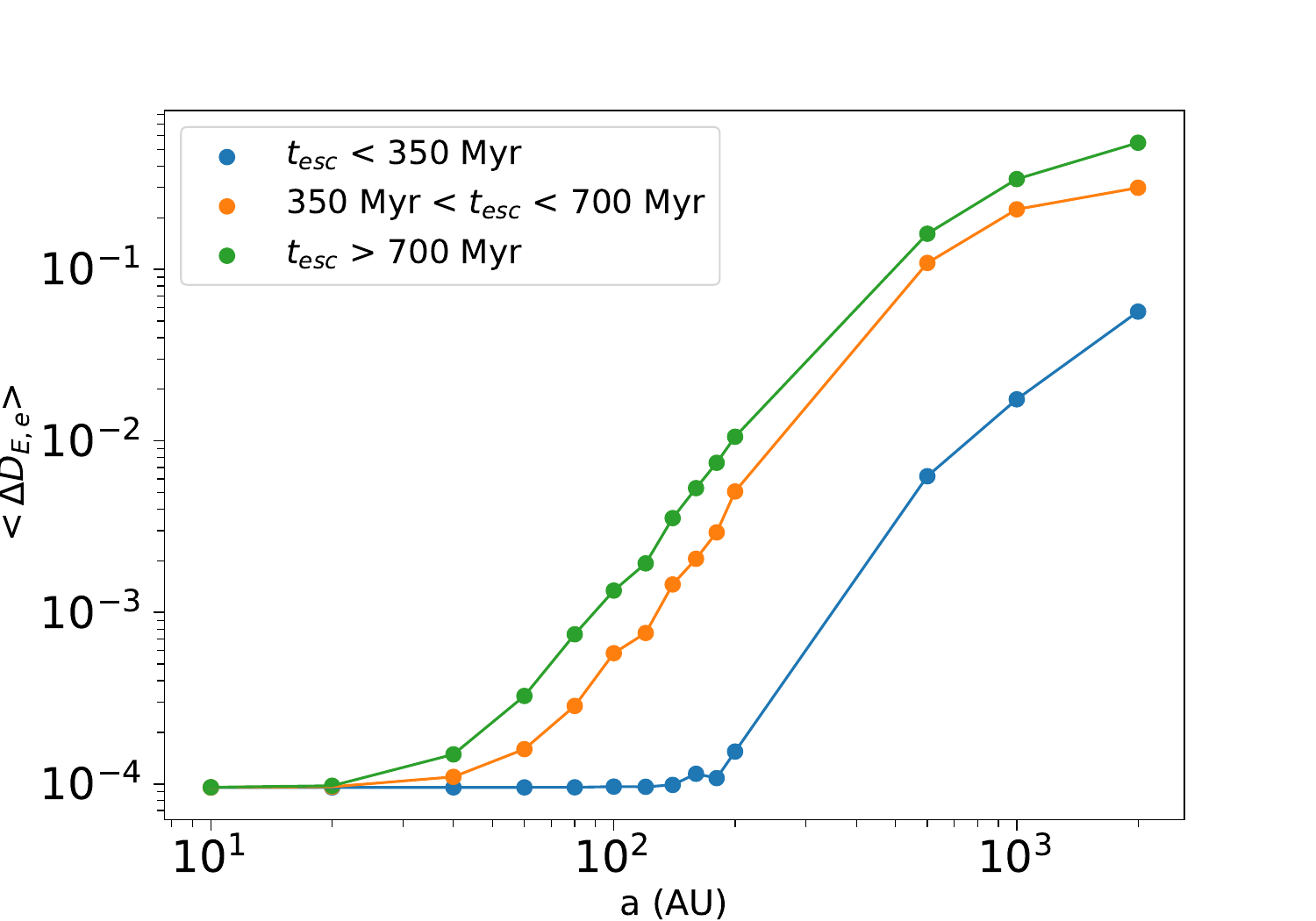}
    \caption{Distribution of each surviving planet's distance in ($E/E_0$,$e$) parameter space from the initial state of (1,0), with the planets initial semi-major axis noted in the panel. Each bar is further broken down to illustrate the contribution of planets from systems with survival fractions less than $33\%$ (green), between $33\%$ and $66\%$ (orange), and above $66\%$ (blue) are shown. While outer planets are more easily perturbed to new orbits than inner planets, regardless of survival fraction, it is possible for inner planets to be significantly perturbed in systems with low and high survival fractions.}
    \label{fig:avg_dist_tesc}
\end{figure}

As a proxy for how strongly a surviving planet has been perturbed from its initial orbit, we calculated its distance in ($E/E_0$,$e-e_0$) parameter space from the initial state of (1,0), which we denote $\Delta D_{E,e}$. This distance should be zero for an unperturbed system and high for cases where planets are heavily perturbed while remaining bound to their host star. Figure \ref{fig:avg_dist_tesc} demonstrates the average $\Delta D_{E,e}$ of planets with the same initial semi-major axis orbiting stars with cluster escape times less than 350 Myr (blue), between 350 Myr and 700 Myr (orange), and more than 700 Myr (green).

Regardless of host star escape time, planets with initial orbital distances less than or equal to 20 au will on average have orbits that are unchanged by the birth cluster environment. For systems that escape the cluster within the first 350 Myr, this statement extends to planets with initial orbital distances less than 200 au. As previously noted, host stars that escape early have initial mean orbital distances greater than 8 pc and therefore experience fewer and weaker perturbations. Of these early escaping systems, only for planets with initial orbital distances greater than 1000 au does the average $\Delta D_{E,e}$ surpass 0.02, which would result in them being outside of the left panel inset in Figure \ref{fig:ecc_energya2}.

Figure \ref{fig:avg_dist_tesc} further demonstrates that for planets with initial orbital distances greater than 20 au, planetary system that remain in the birth cluster longer (and therefore experience a denser cluster environment) are more likely to have planets with perturbed orbits. However for planets with initial orbital distances less than or equal to 100 au, the average perturbation is still small and the average planet would remain in the left panel inset in Figure \ref{fig:ecc_energya2}. Only planets with initial semi-major axes of 600 au or more is the average $\Delta D_{E,e}$ signficant. These are average trends, and as illustrated in Figure \ref{fig:ecc_energya2}, random encounters can still result in a heavily perturbed planets with any initial semi-major axis.

Combining the results of Figures \ref{fig:ecc_energya2} and \ref{fig:avg_dist_tesc} yield two additional conclusions that pertain to our ability to judge a system's dynamical history based on the properties of its surviving planets. First, we see several cases with early escape times where some of the surviving planets have been perturbed to different orbits. Hence observing a heavily perturbed system does not necessarily indicate that it escaped its host cluster late, had a small initial mean orbital distance, or has lost some planets. Second, and similarly, we see cases where outer planets have been stripped entirely from a host star while the remaining inner planets maintain their original orbits. Therefore we can't assume that just because the observed planet's in a system all have circular orbits that the host star did not spend enough time in its birth cluster to experience perturbations that strip outer planets entirely. Hence the orbital properties of surviving planets are not necessarily indicative of the systems birth environment and subsequent evolution.

\section{Conclusions}\label{sec:conclusion}

Motivated by the lack of detected exoplanets in star clusters, we performed direct N-body simulations of planetary systems evolving within a star cluster using the AMUSE framework. We specifically explore how the survival rate of planetary systems in star clusters depends on the a host star's location along the stellar stream that forms out of the star cluster dissolution. The location of a given star along the stream is indicative of its dynamical history within the progenitor cluster, as stream location has been shown to be a proxy for the star's escape time and the state of the cluster when the star escaped \citep{Webb2022}. Since stellar streams have significantly lower densities compared to star clusters, searching for exoplanets along these streams is made easier as there is less overcrowding. Observing the distribution of planetary system properties along a stellar stream will allow for streams to serve as a means to indirectly investigate the evolution of planetary systems within clustered environments.

A detailed analysis of our simulations reveals the following:

\begin{itemize}

\item A star's location along a stellar stream is an excellent proxy of its escape time and dynamical history. Stars near the edge of a stream primarily experienced low-density environments and escaped their host cluster early, while stars near the centre of the stream primarily experienced higher density environments and escaped their host cluster late. \\

\item The probability of a planet's survival depends on the planet's initial semi-major axis $a_0$, the initial mean orbital distance of the host star $\langle r_0 \rangle$, and the host star's escape time. Early escape times and short planet orbital distances increase the probability of survival, while late escape times can yield a wide range of survival fractions. For the model cluster considered here, Equations \ref{eqn:prob}-\ref{eqn:beta} yield a strong quantitative fit of this relationship.\\

\item The degree of orbital perturbation experienced by surviving planets increases with both the planet's initial orbital distance and the time the host star remains in the cluster. \\

\item The orbits of surviving planets in a planetary system are not a strong indicator of the system's dynamical history or survival fraction.

\end{itemize}

Our results suggest that observers searching for exoplanets along streams should focus their efforts on stars located near the stream's edge. In fact, when our model cluster first reaches dissolution, the on-sky distances between stars along the stream are greater than the TESS field of view \footnote{https://tess.mit.edu/science/} when at closest approach (except at the very centre). Hence contamination from other stream stars will not be an issue when searching for transit events. These systems most likely escaped their host cluster early, had larger initial mean orbit distances within the cluster, and have the highest probability of being completely unperturbed. \citet{Curtis2019} have already noted that Pisces–Eridanus is an excellent candidate for finding exoplanets within a stream environment. The fraction of edge stars with planetary systems may also serve as an indication of the fraction of protoplanetary disks that form planets in a star cluster.

In order to apply these results to observed stellar streams, additional simulations are required to explore a range of host cluster masses, sizes, and dissolution times. Additional parameters beyond a star's location along the stream and a planet's semi-major axis may be required to estimate the probability of planet survival along \textit{any} stellar stream. However to first order we already know that denser clusters and clusters with longer dissolution times will be less friendly environments for planetary systems than presented here. The Sun's likely birth cluster, for example, is estimated to be denser and less massive than the model cluster presented in this study \citep{PortegiesZwart2009b, Adams2010}. However, since our model cluster orbits at 5 kpc, their dissolution times are comparable and the Sun's birth cluster can be treated as a high-density analogue of our model cluster. We therefore suspect that the Sun most likely had a large initial mean orbital distance and escaped its birth environment not long after formation given the near-circular and co-planar orbits of its planets. However, an early escape is not \textit{essential} as we observe some planetary systems around Sun-like stars in our simulations with late escape times and planetary systems that are flat with near-circular orbits.

%% Please use the acknowledgment and contribution environments. This will 
%% be anonomyized when the "anonymous" style option is used. 
\begin{acknowledgments}
JW and DM would like to thank Yanqin Wu for helpful discussions about the detectability of exoplanets with TESS. JW also acknowledges computational facilities provided by Jo Bovy and financial support from NSERC (funding reference number RGPIN-2020-04712) and an Ontario Early Researcher Award (ER16-12-061). The Dunlap Institute is funded through an endowment established by the David Dunlap family and the University of Toronto.
\end{acknowledgments}

%% For this sample we use BibTeX plus aasjournalv7.bst to generate the

\bibliography{sample701}{}

@BOOK{Heggie2003,
       author = {{Heggie}, Douglas and {Hut}, Piet},
        title = "{The Gravitational Million-Body Problem: A Multidisciplinary Approach to Star Cluster Dynamics}",
         year = 2003,
       adsurl = {https://ui.adsabs.harvard.edu/abs/2003gmbp.book.....H}
}

@ARTICLE{Pelupessy2013,
       author = {{Pelupessy}, F.~I. and {van Elteren}, A. and {de Vries}, N. and {McMillan}, S.~L.~W. and {Drost}, N. and {Portegies Zwart}, S.~F.},
        title = "{The Astrophysical Multipurpose Software Environment}",
      journal = {\aap},
         year = 2013,
        month = sep,
       volume = {557},
          eid = {A84},
        pages = {A84},
          doi = {10.1051/0004-6361/201321252},
archivePrefix = {arXiv},
       eprint = {1307.3016},
 primaryClass = {astro-ph.IM},
       adsurl = {https://ui.adsabs.harvard.edu/abs/2013A&A...557A..84P}
}

@ARTICLE{Arzoumanian2023,
       author = {{Arzoumanian}, Doris and {Arakawa}, Sota and {Kobayashi}, Masato I.~N. and {Iwasaki}, Kazunari and {Fukuda}, Kohei and {Mori}, Shoji and {Hirai}, Yutaka and {Kunitomo}, Masanobu and {Kumar}, M.~S. Nanda and {Kokubo}, Eiichiro},
        title = "{Insights on the Sun Birth Environment in the Context of Star Cluster Formation in Hub-Filament Systems}",
      journal = {\apjl},
         year = 2023,
        month = apr,
       volume = {947},
       number = {2},
          eid = {L29},
        pages = {L29},
          doi = {10.3847/2041-8213/acc849},
archivePrefix = {arXiv},
       eprint = {2303.15695},
 primaryClass = {astro-ph.GA},
       adsurl = {https://ui.adsabs.harvard.edu/abs/2023ApJ...947L..29A}
}

@ARTICLE{Wang2015,
       author = {{Wang}, Long and {Spurzem}, Rainer and {Aarseth}, Sverre and {Nitadori}, Keigo and {Berczik}, Peter and {Kouwenhoven}, M.~B.~N. and {Naab}, Thorsten},
        title = "{NBODY6++GPU: ready for the gravitational million-body problem}",
      journal = {\mnras},
         year = 2015,
        month = jul,
       volume = {450},
       number = {4},
        pages = {4070-4080},
          doi = {10.1093/mnras/stv817},
archivePrefix = {arXiv},
       eprint = {1504.03687},
 primaryClass = {astro-ph.IM},
       adsurl = {https://ui.adsabs.harvard.edu/abs/2015MNRAS.450.4070W}
}

@ARTICLE{Cai2019,
       author = {{Cai}, Maxwell X. and {Portegies Zwart}, S. and {Kouwenhoven}, M.~B.~N. and {Spurzem}, Rainer},
        title = "{On the survivability of planets in young massive clusters and its implication of planet orbital architectures in globular clusters}",
      journal = {\mnras},
         year = 2019,
        month = nov,
       volume = {489},
       number = {3},
        pages = {4311-4321},
          doi = {10.1093/mnras/stz2467},
archivePrefix = {arXiv},
       eprint = {1903.02316},
 primaryClass = {astro-ph.EP},
       adsurl = {https://ui.adsabs.harvard.edu/abs/2019MNRAS.489.4311C}
}

@ARTICLE{Rein2012,
       author = {{Rein}, H. and {Liu}, S. -F.},
        title = "{REBOUND: an open-source multi-purpose N-body code for collisional dynamics}",
      journal = {\aap},
         year = 2012,
        month = jan,
       volume = {537},
          eid = {A128},
        pages = {A128},
          doi = {10.1051/0004-6361/201118085},
archivePrefix = {arXiv},
       eprint = {1110.4876},
 primaryClass = {astro-ph.EP},
       adsurl = {https://ui.adsabs.harvard.edu/abs/2012A&A...537A.128R}
}

@ARTICLE{Bovy2015,
       author = {{Bovy}, Jo},
        title = "{galpy: A python Library for Galactic Dynamics}",
      journal = {\apjs},
         year = 2015,
        month = feb,
       volume = {216},
       number = {2},
          eid = {29},
        pages = {29},
          doi = {10.1088/0067-0049/216/2/29},
archivePrefix = {arXiv},
       eprint = {1412.3451},
 primaryClass = {astro-ph.GA},
       adsurl = {https://ui.adsabs.harvard.edu/abs/2015ApJS..216...29B}
}

@ARTICLE{Adams2010,
       author = {{Adams}, Fred C.},
        title = "{The Birth Environment of the Solar System}",
      journal = {\araa},
         year = 2010,
        month = sep,
       volume = {48},
        pages = {47-85},
          doi = {10.1146/annurev-astro-081309-130830},
archivePrefix = {arXiv},
       eprint = {1001.5444},
 primaryClass = {astro-ph.SR},
       adsurl = {https://ui.adsabs.harvard.edu/abs/2010ARA&A..48...47A}
}

@ARTICLE{Winn2015,
       author = {{Winn}, Joshua N. and {Fabrycky}, Daniel C.},
        title = "{The Occurrence and Architecture of Exoplanetary Systems}",
      journal = {\araa},
         year = 2015,
        month = aug,
       volume = {53},
        pages = {409-447},
          doi = {10.1146/annurev-astro-082214-122246},
archivePrefix = {arXiv},
       eprint = {1410.4199},
 primaryClass = {astro-ph.EP},
       adsurl = {https://ui.adsabs.harvard.edu/abs/2015ARA&A..53..409W}
}

@BOOK{Aarseth2003,
       author = {{Aarseth}, Sverre J.},
        title = "{Gravitational N-Body Simulations}",
         year = 2003,
         publisher = {Cambridge University Press} ,
       adsurl = {https://ui.adsabs.harvard.edu/abs/2003gnbs.book.....A}
}

@ARTICLE{Lada2003,
       author = {{Lada}, Charles J. and {Lada}, Elizabeth A.},
        title = "{Embedded Clusters in Molecular Clouds}",
      journal = {\araa},
         year = 2003,
        month = jan,
       volume = {41},
        pages = {57-115},
          doi = {10.1146/annurev.astro.41.011802.094844},
archivePrefix = {arXiv},
       eprint = {astro-ph/0301540},
 primaryClass = {astro-ph},
       adsurl = {https://ui.adsabs.harvard.edu/abs/2003ARA&A..41...57L}
}

@ARTICLE{Spitzer1958,
       author = {{Spitzer}, Lyman, Jr.},
        title = "{Disruption of Galactic Clusters.}",
      journal = {\apj},
         year = 1958,
        month = jan,
       volume = {127},
        pages = {17},
          doi = {10.1086/146435},
       adsurl = {https://ui.adsabs.harvard.edu/abs/1958ApJ...127...17S}
}

@BOOK{Spitzer1987,
       author = {{Spitzer}, Lyman},
        title = "{Dynamical evolution of globular clusters}",
         year = 1987,
       adsurl = {https://ui.adsabs.harvard.edu/abs/1987degc.book.....S}
}

@ARTICLE{Cassan2012,
       author = {{Cassan}, A. and {Kubas}, D. and {Beaulieu}, J. -P. and {Dominik}, M. and {Horne}, K. and {Greenhill}, J. and {Wambsganss}, J. and {Menzies}, J. and {Williams}, A. and {J{\o}rgensen}, U.~G. and {Udalski}, A. and {Bennett}, D.~P. and {Albrow}, M.~D. and {Batista}, V. and {Brillant}, S. and {Caldwell}, J.~A.~R. and {Cole}, A. and {Coutures}, Ch. and {Cook}, K.~H. and {Dieters}, S. and {Dominis Prester}, D. and {Donatowicz}, J. and {Fouqu{\'e}}, P. and {Hill}, K. and {Kains}, N. and {Kane}, S. and {Marquette}, J. -B. and {Martin}, R. and {Pollard}, K.~R. and {Sahu}, K.~C. and {Vinter}, C. and {Warren}, D. and {Watson}, B. and {Zub}, M. and {Sumi}, T. and {Szyma{\'n}ski}, M.~K. and {Kubiak}, M. and {Poleski}, R. and {Soszynski}, I. and {Ulaczyk}, K. and {Pietrzy{\'n}ski}, G. and {Wyrzykowski}, {\L}.},
        title = "{One or more bound planets per Milky Way star from microlensing observations}",
      journal = {\nat},
         year = 2012,
        month = jan,
       volume = {481},
       number = {7380},
        pages = {167-169},
          doi = {10.1038/nature10684},
archivePrefix = {arXiv},
       eprint = {1202.0903},
 primaryClass = {astro-ph.EP},
       adsurl = {https://ui.adsabs.harvard.edu/abs/2012Natur.481..167C}
}

@ARTICLE{Backer1993,
       author = {{Backer}, D.~C. and {Foster}, R.~S. and {Sallmen}, S.},
        title = "{A second companion of the millisecond pulsar 1620 - 26}",
      journal = {\nat},
         year = 1993,
        month = oct,
       volume = {365},
       number = {6449},
        pages = {817-819},
          doi = {10.1038/365817a0},
       adsurl = {https://ui.adsabs.harvard.edu/abs/1993Natur.365..817B}
}

@ARTICLE{Webb2022,
       author = {{Webb}, Jeremy J. and {Bovy}, Jo},
        title = "{Variation in the stellar mass function along stellar streams}",
      journal = {\mnras},
         year = 2022,
        month = feb,
       volume = {510},
       number = {1},
        pages = {774-785},
          doi = {10.1093/mnras/stab3451},
archivePrefix = {arXiv},
       eprint = {2108.02217},
 primaryClass = {astro-ph.GA},
       adsurl = {https://ui.adsabs.harvard.edu/abs/2022MNRAS.510..774W}
}

@ARTICLE{Pang2021,
       author = {{Pang}, Xiaoying and {Li}, Yuqian and {Yu}, Zeqiu and {Tang}, Shih-Yun and {Dinnbier}, Franti{\v{s}}ek and {Kroupa}, Pavel and {Pasquato}, Mario and {Kouwenhoven}, M.~B.~N.},
        title = "{3D Morphology of Open Clusters in the Solar Neighborhood with Gaia EDR 3: Its Relation to Cluster Dynamics}",
      journal = {\apj},
         year = 2021,
        month = may,
       volume = {912},
       number = {2},
          eid = {162},
        pages = {162},
          doi = {10.3847/1538-4357/abeaac},
archivePrefix = {arXiv},
       eprint = {2102.10508},
 primaryClass = {astro-ph.GA},
       adsurl = {https://ui.adsabs.harvard.edu/abs/2021ApJ...912..162P}
}

@ARTICLE{Ye2021,
       author = {{Ye}, Xianhao and {Zhao}, Jinkun and {Zhang}, Jiajun and {Yang}, Yong and {Zhao}, Gang},
        title = "{Extended Tidal Tails of IC 4756 Detected by Gaia EDR3}",
      journal = {\aj},
         year = 2021,
        month = oct,
       volume = {162},
       number = {4},
          eid = {171},
        pages = {171},
          doi = {10.3847/1538-3881/ac1f1f},
archivePrefix = {arXiv},
       eprint = {2110.08104},
 primaryClass = {astro-ph.GA},
       adsurl = {https://ui.adsabs.harvard.edu/abs/2021AJ....162..171Y}
}

@ARTICLE{Newton2021,
       author = {{Newton}, Elisabeth R. and {Mann}, Andrew W. and {Kraus}, Adam L. and {Livingston}, John H. and {Vanderburg}, Andrew and {Curtis}, Jason L. and {Thao}, Pa Chia and {Hawkins}, Keith and {Wood}, Mackenna L. and {Rizzuto}, Aaron C. and {Soubkiou}, Abderahmane and {Tofflemire}, Benjamin M. and {Zhou}, George and {Crossfield}, Ian J.~M. and {Pearce}, Logan A. and {Collins}, Karen A. and {Conti}, Dennis M. and {Tan}, Thiam-Guan and {Villeneuva}, Steven and {Spencer}, Alton and {Dragomir}, Diana and {Quinn}, Samuel N. and {Jensen}, Eric L.~N. and {Collins}, Kevin I. and {Stockdale}, Chris and {Cloutier}, Ryan and {Hellier}, Coel and {Benkhaldoun}, Zouhair and {Ziegler}, Carl and {Brice{\~n}o}, C{\'e}sar and {Law}, Nicholas and {Benneke}, Bj{\"o}rn and {Christiansen}, Jessie L. and {Gorjian}, Varoujan and {Kane}, Stephen R. and {Kreidberg}, Laura and {Morales}, Farisa Y. and {Werner}, Michael W. and {Twicken}, Joseph D. and {Levine}, Alan M. and {Ciardi}, David R. and {Guerrero}, Natalia M. and {Hesse}, Katharine and {Quintana}, Elisa V. and {Shiao}, Bernie and {Smith}, Jeffrey C. and {Torres}, Guillermo and {Ricker}, George R. and {Vanderspek}, Roland and {Seager}, Sara and {Winn}, Joshua N. and {Jenkins}, Jon M. and {Latham}, David W.},
        title = "{TESS Hunt for Young and Maturing Exoplanets (THYME). IV. Three Small Planets Orbiting a 120 Myr Old Star in the Pisces-Eridanus Stream}",
      journal = {\aj},
         year = 2021,
        month = feb,
       volume = {161},
       number = {2},
          eid = {65},
        pages = {65},
          doi = {10.3847/1538-3881/abccc6},
archivePrefix = {arXiv},
       eprint = {2102.06049},
 primaryClass = {astro-ph.EP},
       adsurl = {https://ui.adsabs.harvard.edu/abs/2021AJ....161...65N}
}

@ARTICLE{Plummer1911,
       author = {{Plummer}, H.~C.},
        title = "{On the problem of distribution in globular star clusters}",
      journal = {\mnras},
         year = 1911,
        month = mar,
       volume = {71},
        pages = {460-470},
          doi = {10.1093/mnras/71.5.460},
       adsurl = {https://ui.adsabs.harvard.edu/abs/1911MNRAS..71..460P}
}

@ARTICLE{Kroupa2001,
       author = {{Kroupa}, Pavel},
        title = "{On the variation of the initial mass function}",
      journal = {\mnras},
         year = 2001,
        month = apr,
       volume = {322},
       number = {2},
        pages = {231-246},
          doi = {10.1046/j.1365-8711.2001.04022.x},
archivePrefix = {arXiv},
       eprint = {astro-ph/0009005},
 primaryClass = {astro-ph},
       adsurl = {https://ui.adsabs.harvard.edu/abs/2001MNRAS.322..231K}
}

@ARTICLE{Rein2015,
       author = {{Rein}, Hanno and {Spiegel}, David S.},
        title = "{IAS15: a fast, adaptive, high-order integrator for gravitational dynamics, accurate to machine precision over a billion orbits}",
      journal = {\mnras},
         year = 2015,
        month = jan,
       volume = {446},
       number = {2},
        pages = {1424-1437},
          doi = {10.1093/mnras/stu2164},
archivePrefix = {arXiv},
       eprint = {1409.4779},
 primaryClass = {astro-ph.EP},
       adsurl = {https://ui.adsabs.harvard.edu/abs/2015MNRAS.446.1424R}
}

@ARTICLE{Drass2016,
       author = {{Drass}, H. and {Haas}, M. and {Chini}, R. and {Bayo}, A. and {Hackstein}, M. and {Hoffmeister}, V. and {Godoy}, N. and {Vogt}, N.},
        title = "{The bimodal initial mass function in the Orion nebula cloud}",
      journal = {\mnras},
         year = 2016,
        month = sep,
       volume = {461},
       number = {2},
        pages = {1734-1744},
          doi = {10.1093/mnras/stw1094},
archivePrefix = {arXiv},
       eprint = {1605.03600},
 primaryClass = {astro-ph.GA},
       adsurl = {https://ui.adsabs.harvard.edu/abs/2016MNRAS.461.1734D}
}

@ARTICLE{Batista2014,
       author = {{Batista}, S.~F.~A. and {Adibekyan}, V. Zh. and {Sousa}, S.~G. and {Santos}, N.~C. and {Delgado Mena}, E. and {Hakobyan}, A.~A.},
        title = "{Searching for solar siblings among the HARPS data}",
      journal = {\aap},
         year = 2014,
        month = apr,
       volume = {564},
          eid = {A43},
        pages = {A43},
          doi = {10.1051/0004-6361/201423645},
archivePrefix = {arXiv},
       eprint = {1403.1506},
 primaryClass = {astro-ph.GA},
       adsurl = {https://ui.adsabs.harvard.edu/abs/2014A&A...564A..43B}
}

@ARTICLE{Pfalzner2020,
       author = {{Pfalzner}, Susanne and {Vincke}, Kirsten},
        title = "{Cradle(s) of the Sun}",
      journal = {\apj},
         year = 2020,
        month = jul,
       volume = {897},
       number = {1},
          eid = {60},
        pages = {60},
          doi = {10.3847/1538-4357/ab9533},
archivePrefix = {arXiv},
       eprint = {2005.11260},
 primaryClass = {astro-ph.EP},
       adsurl = {https://ui.adsabs.harvard.edu/abs/2020ApJ...897...60P}
}

@ARTICLE{Batista2012,
       author = {{Batista}, S{\'e}rgio Filipe Assun{\c{c}}{\~a}o and {Fernandes}, Jo{\~a}o},
        title = "{Lost siblings of the Sun: Revisiting the FGK potential candidates}",
      journal = {\na},
         year = 2012,
        month = jul,
       volume = {17},
       number = {5},
        pages = {514-519},
          doi = {10.1016/j.newast.2011.12.001},
       adsurl = {https://ui.adsabs.harvard.edu/abs/2012NewA...17..514B}
}

@ARTICLE{Adibekyan2018,
       author = {{Adibekyan}, V. and {de Laverny}, P. and {Recio-Blanco}, A. and {Sousa}, S.~G. and {Delgado-Mena}, E. and {Kordopatis}, G. and {Ferreira}, A.~C.~S. and {Santos}, N.~C. and {Hakobyan}, A.~A. and {Tsantaki}, M.},
        title = "{The AMBRE project: searching for the closest solar siblings}",
      journal = {\aap},
         year = 2018,
        month = nov,
       volume = {619},
          eid = {A130},
        pages = {A130},
          doi = {10.1051/0004-6361/201834285},
archivePrefix = {arXiv},
       eprint = {1810.01813},
 primaryClass = {astro-ph.SR},
       adsurl = {https://ui.adsabs.harvard.edu/abs/2018A&A...619A.130A}
}

@ARTICLE{Brown2010,
       author = {{Brown}, Anthony G.~A. and {Portegies Zwart}, Simon F. and {Bean}, Jennifer},
        title = "{The quest for the Sun's siblings: an exploratory search in the Hipparcos Catalogue}",
      journal = {\mnras},
         year = 2010,
        month = sep,
       volume = {407},
       number = {1},
        pages = {458-464},
          doi = {10.1111/j.1365-2966.2010.16921.x},
archivePrefix = {arXiv},
       eprint = {1004.4284},
 primaryClass = {astro-ph.GA},
       adsurl = {https://ui.adsabs.harvard.edu/abs/2010MNRAS.407..458B}
}

@ARTICLE{Webb2020,
       author = {{Webb}, Jeremy J. and {Price-Jones}, Natalie and {Bovy}, Jo and {Portegies Zwart}, Simon and {Hunt}, Jason A.~S. and {Mackereth}, J. Ted and {Leung}, Henry W.},
        title = "{Searching for solar siblings in APOGEE and Gaia DR2 with N-body simulations}",
      journal = {\mnras},
         year = 2020,
        month = may,
       volume = {494},
       number = {2},
        pages = {2268-2279},
          doi = {10.1093/mnras/staa788},
archivePrefix = {arXiv},
       eprint = {1910.01646},
 primaryClass = {astro-ph.GA},
       adsurl = {https://ui.adsabs.harvard.edu/abs/2020MNRAS.494.2268W}
}

@ARTICLE{PortegiesZwart2009b,
       author = {{Portegies Zwart}, Simon F.},
        title = "{The Lost Siblings of the Sun}",
      journal = {\apjl},
         year = 2009,
        month = may,
       volume = {696},
       number = {1},
        pages = {L13-L16},
          doi = {10.1088/0004-637X/696/1/L13},
archivePrefix = {arXiv},
       eprint = {0903.0237},
 primaryClass = {astro-ph.GA},
       adsurl = {https://ui.adsabs.harvard.edu/abs/2009ApJ...696L..13P}
}

@ARTICLE{PortegiesZwart2009,
       author = {{Portegies Zwart}, Simon and {McMillan}, Steve and {Harfst}, Stefan and {Groen}, Derek and {Fujii}, Michiko and {Nuall{\'a}in}, Breannd{\'a}n {\'O}. and {Glebbeek}, Evert and {Heggie}, Douglas and {Lombardi}, James and {Hut}, Piet and {Angelou}, Vangelis and {Banerjee}, Sambaran and {Belkus}, Houria and {Fragos}, Tassos and {Fregeau}, John and {Gaburov}, Evghenii and {Izzard}, Rob and {Juri{\'c}}, Mario and {Justham}, Stephen and {Sottoriva}, Andrea and {Teuben}, Peter and {van Bever}, Joris and {Yaron}, Ofer and {Zemp}, Marcel},
        title = "{A multiphysics and multiscale software environment for modeling astrophysical systems}",
      journal = {\na},
         year = 2009,
        month = may,
       volume = {14},
       number = {4},
        pages = {369-378},
          doi = {10.1016/j.newast.2008.10.006},
archivePrefix = {arXiv},
       eprint = {0807.1996},
 primaryClass = {astro-ph},
       adsurl = {https://ui.adsabs.harvard.edu/abs/2009NewA...14..369P}
}

@ARTICLE{PortegiesZwart2013,
       author = {{Portegies Zwart}, S. and {McMillan}, S.~L.~W. and {van Elteren}, E. and {Pelupessy}, I. and {de Vries}, N.},
        title = "{Multi-physics simulations using a hierarchical interchangeable software interface}",
      journal = {Computer Physics Communications},
         year = 2013,
        month = mar,
       volume = {184},
       number = {3},
        pages = {456-468},
          doi = {10.1016/j.cpc.2012.09.024},
archivePrefix = {arXiv},
       eprint = {1204.5522},
 primaryClass = {astro-ph.IM},
       adsurl = {https://ui.adsabs.harvard.edu/abs/2013CoPhC.184..456P}
}

@ARTICLE{2018MNRAS.474.5114C,
       author = {{Cai}, Maxwell Xu and {Portegies Zwart}, Simon and {van Elteren}, Arjen},
        title = "{The signatures of the parental cluster on field planetary systems}",
      journal = {\mnras},
         year = 2018,
        month = mar,
       volume = {474},
       number = {4},
        pages = {5114-5121},
          doi = {10.1093/mnras/stx3064},
archivePrefix = {arXiv},
       eprint = {1711.01274},
 primaryClass = {astro-ph.EP},
       adsurl = {https://ui.adsabs.harvard.edu/abs/2018MNRAS.474.5114C}
}

@ARTICLE{2017MNRAS.470.4337C,
       author = {{Cai}, Maxwell Xu and {Kouwenhoven}, M.~B.~N. and {Portegies Zwart}, Simon F. and {Spurzem}, Rainer},
        title = "{Stability of multiplanetary systems in star clusters}",
      journal = {\mnras},
         year = 2017,
        month = oct,
       volume = {470},
       number = {4},
        pages = {4337-4353},
          doi = {10.1093/mnras/stx1464},
archivePrefix = {arXiv},
       eprint = {1706.03789},
 primaryClass = {astro-ph.EP},
       adsurl = {https://ui.adsabs.harvard.edu/abs/2017MNRAS.470.4337C}
}

@ARTICLE{2015ApJS..219...31C,
       author = {{Cai}, Maxwell Xu and {Meiron}, Yohai and {Kouwenhoven}, M.~B.~N. and {Assmann}, Paulina and {Spurzem}, Rainer},
        title = "{Block Time Step Storage Scheme for Astrophysical N-body Simulations}",
      journal = {\apjs},
         year = 2015,
        month = aug,
       volume = {219},
       number = {2},
          eid = {31},
        pages = {31},
          doi = {10.1088/0067-0049/219/2/31},
archivePrefix = {arXiv},
       eprint = {1506.07591},
 primaryClass = {astro-ph.IM},
       adsurl = {https://ui.adsabs.harvard.edu/abs/2015ApJS..219...31C}
}

@ARTICLE{PortegiesZwart2010,
       author = {{Portegies Zwart}, Simon F. and {McMillan}, Stephen L.~W. and {Gieles}, Mark},
        title = "{Young Massive Star Clusters}",
      journal = {\araa},
         year = 2010,
        month = sep,
       volume = {48},
        pages = {431-493},
          doi = {10.1146/annurev-astro-081309-130834},
archivePrefix = {arXiv},
       eprint = {1002.1961},
 primaryClass = {astro-ph.GA},
       adsurl = {https://ui.adsabs.harvard.edu/abs/2010ARA&A..48..431P}
}

@ARTICLE{PortegiesZwart2021,
       author = {{Portegies Zwart}, S.},
        title = "{Oort cloud Ecology. I. Extra-solar Oort clouds and the origin of asteroidal interlopers}",
      journal = {\aap},
         year = 2021,
        month = mar,
       volume = {647},
          eid = {A136},
        pages = {A136},
          doi = {10.1051/0004-6361/202038888},
archivePrefix = {arXiv},
       eprint = {2011.08257},
 primaryClass = {astro-ph.EP},
       adsurl = {https://ui.adsabs.harvard.edu/abs/2021A&A...647A.136P}
}

@ARTICLE{Hanse2018,
       author = {{Hanse}, J. and {J{\'\i}lkov{\'a}}, L. and {Portegies Zwart}, S.~F. and {Pelupessy}, F.~I.},
        title = "{Capture of exocomets and the erosion of the Oort cloud due to stellar encounters in the Galaxy}",
      journal = {\mnras},
         year = 2018,
        month = feb,
       volume = {473},
       number = {4},
        pages = {5432-5445},
          doi = {10.1093/mnras/stx2721},
archivePrefix = {arXiv},
       eprint = {1710.05866},
 primaryClass = {astro-ph.EP},
       adsurl = {https://ui.adsabs.harvard.edu/abs/2018MNRAS.473.5432H}
}

@ARTICLE{vanElteren2019,
       author = {{van Elteren}, A. and {Portegies Zwart}, S. and {Pelupessy}, I. and {Cai}, M.~X. and {McMillan}, S.~L.~W.},
        title = "{Survivability of planetary systems in young and dense star clusters}",
      journal = {\aap},
         year = 2019,
        month = apr,
       volume = {624},
          eid = {A120},
        pages = {A120},
          doi = {10.1051/0004-6361/201834641},
archivePrefix = {arXiv},
       eprint = {1902.04652},
 primaryClass = {astro-ph.SR},
       adsurl = {https://ui.adsabs.harvard.edu/abs/2019A&A...624A.120V}
}

@ARTICLE{DaffernPowell2022,
       author = {{Daffern-Powell}, Emma C. and {Parker}, Richard J. and {Quanz}, Sascha P.},
        title = "{The Great Planetary Heist: theft and capture in star-forming regions}",
      journal = {\mnras},
         year = 2022,
        month = jul,
       volume = {514},
       number = {1},
        pages = {920-934},
          doi = {10.1093/mnras/stac1392},
archivePrefix = {arXiv},
       eprint = {2205.07895},
 primaryClass = {astro-ph.EP},
       adsurl = {https://ui.adsabs.harvard.edu/abs/2022MNRAS.514..920D}
}

@ARTICLE{Dai2023,
       author = {{Dai}, Yuan-Zhe and {Liu}, Hui-Gen and {Yang}, Jia-Yi and {Zhou}, Ji-Lin},
        title = "{Understanding the Planetary Formation and Evolution in Star Clusters (UPiC). I. Evidence of Hot Giant Exoplanets Formation Timescales}",
      journal = {\aj},
         year = 2023,
        month = dec,
       volume = {166},
       number = {6},
          eid = {219},
        pages = {219},
          doi = {10.3847/1538-3881/acff67},
archivePrefix = {arXiv},
       eprint = {2306.02610},
 primaryClass = {astro-ph.EP},
       adsurl = {https://ui.adsabs.harvard.edu/abs/2023AJ....166..219D}
}

@ARTICLE{Martin2025,
       author = {{Mart{\'\i}n}, E.~L. and {{\v{Z}}erjal}, M. and {Bouy}, H. and {Martin-Gonzalez}, D. and {Mu{\~n}oz Torres}, S. and {Barrado}, D. and {Olivares}, J. and {P{\'e}rez-Garrido}, A. and {Mas-Buitrago}, P. and {Cruz}, P. and {Solano}, E. and {Zapatero Osorio}, M.~R. and {Lodieu}, N. and {B{\'e}jar}, V.~J.~S. and {Zhang}, J. -Y. and {del Burgo}, C. and {Hu{\'e}lamo}, N. and {Laureijs}, R. and {Mora}, A. and {Saifollahi}, T. and {Cuillandre}, J. -C. and {Schirmer}, M. and {Tata}, R. and {Points}, S. and {Phan-Bao}, N. and {Goldman}, B. and {Casewell}, S.~L. and {Reyl{\'e}}, C. and {Smart}, R.~L. and {Dominguez-Tagle}, C. and {Escobar}, A. and {Sedighi}, N. and {Tsilia}, S. and {Vitas}, N. and {Ayadi}, A. and {Aghanim}, N. and {Altieri}, B. and {Andreon}, S. and {Auricchio}, N. and {Baldi}, M. and {Balestra}, A. and {Bardelli}, S. and {Basset}, A. and {Bender}, R. and {Bonino}, D. and {Branchini}, E. and {Brescia}, M. and {Brinchmann}, J. and {Camera}, S. and {Capobianco}, V. and {Carbone}, C. and {Carretero}, J. and {Casas}, S. and {Castellano}, M. and {Cavuoti}, S. and {Cimatti}, A. and {Congedo}, G. and {Conselice}, C.~J. and {Conversi}, L. and {Copin}, Y. and {Corcione}, L. and {Courbin}, F. and {Courtois}, H.~M. and {Cropper}, M. and {Da Silva}, A. and {Degaudenzi}, H. and {Di Giorgio}, A.~M. and {Dinis}, J. and {Dubath}, F. and {Dupac}, X. and {Dusini}, S. and {Ealet}, A. and {Farina}, M. and {Farrens}, S. and {Ferriol}, S. and {Fosalba}, P. and {Frailis}, M. and {Franceschi}, E. and {Fumana}, M. and {Galeotta}, S. and {Garilli}, B. and {Gillard}, W. and {Gillis}, B. and {Giocoli}, C. and {G{\'o}mez-Alvarez}, P. and {Grazian}, A. and {Grupp}, F. and {Guzzo}, L. and {Haugan}, S.~V.~H. and {Hoar}, J. and {Hoekstra}, H. and {Holmes}, W. and {Hook}, I. and {Hormuth}, F. and {Hornstrup}, A. and {Hu}, D. and {Hudelot}, P. and {Jahnke}, K. and {Jhabvala}, M. and {Keih{\"a}nen}, E. and {Kermiche}, S. and {Kiessling}, A. and {Kilbinger}, M. and {Kitching}, T. and {Kohley}, R. and {Kubik}, B. and {K{\"u}mmel}, M. and {Kunz}, M. and {Kurki-Suonio}, H. and {Le Mignant}, D. and {Ligori}, S. and {Lilje}, P.~B. and {Lindholm}, V. and {Lloro}, I. and {Maino}, D. and {Maiorano}, E. and {Mansutti}, O. and {Marggraf}, O. and {Martinet}, N. and {Marulli}, F. and {Massey}, R. and {Medinaceli}, E. and {Mei}, S. and {Melchior}, M. and {Mellier}, Y. and {Meneghetti}, M. and {Meylan}, G. and {Mohr}, J.~J. and {Moresco}, M. and {Moscardini}, L. and {Niemi}, S. -M. and {Padilla}, C. and {Paltani}, S. and {Pasian}, F. and {Pedersen}, K. and {Percival}, W.~J. and {Pettorino}, V. and {Pires}, S. and {Polenta}, G. and {Poncet}, M. and {Popa}, L.~A. and {Pozzetti}, L. and {Racca}, G.~D. and {Raison}, F. and {Rebolo}, R. and {Renzi}, A. and {Rhodes}, J. and {Riccio}, G. and {Rix}, Hans-Walter and {Romelli}, E. and {Roncarelli}, M. and {Rossetti}, E. and {Saglia}, R. and {Sapone}, D. and {Sartoris}, B. and {Sauvage}, M. and {Scaramella}, R. and {Schneider}, P. and {Secroun}, A. and {Seidel}, G. and {Seiffert}, M. and {Serrano}, S. and {Sirignano}, C. and {Sirri}, G. and {Stanco}, L. and {Tallada-Cresp{\'\i}}, P. and {Taylor}, A.~N. and {Teplitz}, H.~I. and {Tereno}, I. and {Toledo-Moreo}, R. and {Tsyganov}, A. and {Tutusaus}, I. and {Valenziano}, L. and {Vassallo}, T. and {Verdoes Kleijn}, G. and {Wang}, Y. and {Weller}, J. and {Williams}, O.~R. and {Zucca}, E. and {Baccigalupi}, C. and {Willis}, G. and {Simon}, P. and {Mart{\'\i}n-Fleitas}, J. and {Scott}, D.},
        title = "{Euclid: Early Release Observations {\textendash} A glance at free-floating newborn planets in the {\ensuremath{\sigma}} Orionis cluster}",
      journal = {\aap},
         year = 2025,
        month = may,
       volume = {697},
          eid = {A7},
        pages = {A7},
          doi = {10.1051/0004-6361/202450793},
archivePrefix = {arXiv},
       eprint = {2405.13497},
 primaryClass = {astro-ph.EP},
       adsurl = {https://ui.adsabs.harvard.edu/abs/2025A&A...697A...7M}
}

@ARTICLE{Luhman07,
       author = {{Luhman}, K.~L.},
        title = "{The Stellar Population of the Chamaeleon I Star-forming Region}",
      journal = {\apjs},
         year = 2007,
        month = nov,
       volume = {173},
       number = {1},
        pages = {104-136},
          doi = {10.1086/520114},
archivePrefix = {arXiv},
       eprint = {0710.3037},
 primaryClass = {astro-ph},
       adsurl = {https://ui.adsabs.harvard.edu/abs/2007ApJS..173..104L}
}

@ARTICLE{Scholz2012,
       author = {{Scholz}, Alexander and {Jayawardhana}, Ray and {Muzic}, Koraljka and {Geers}, Vincent and {Tamura}, Motohide and {Tanaka}, Ichi},
        title = "{Substellar Objects in Nearby Young Clusters (SONYC). VI. The Planetary-mass Domain of NGC 1333}",
      journal = {\apj},
         year = 2012,
        month = sep,
       volume = {756},
       number = {1},
          eid = {24},
        pages = {24},
          doi = {10.1088/0004-637X/756/1/24},
archivePrefix = {arXiv},
       eprint = {1207.1449},
 primaryClass = {astro-ph.SR},
       adsurl = {https://ui.adsabs.harvard.edu/abs/2012ApJ...756...24S}
}

@INPROCEEDINGS{Sigurdsson1993,
       author = {{Sigurdsson}, S.},
        title = "{Planets in globular clusters?}",
    booktitle = {Planets Around Pulsars},
         year = 1993,
       editor = {{Phillips}, J.~A. and {Thorsett}, Steve E. and {Kulkarni}, Shri R.},
       series = {Astronomical Society of the Pacific Conference Series},
       volume = {36},
        month = jan,
        pages = {173-179},
       adsurl = {https://ui.adsabs.harvard.edu/abs/1993ASPC...36..173S}
}

@ARTICLE{Curtis2019,
       author = {{Curtis}, Jason L. and {Ag{\"u}eros}, Marcel A. and {Mamajek}, Eric E. and {Wright}, Jason T. and {Cummings}, Jeffrey D.},
        title = "{TESS Reveals that the Nearby Pisces-Eridanus Stellar Stream is only 120 Myr Old}",
      journal = {\aj},
         year = 2019,
        month = aug,
       volume = {158},
       number = {2},
          eid = {77},
        pages = {77},
          doi = {10.3847/1538-3881/ab2899},
archivePrefix = {arXiv},
       eprint = {1905.10588},
 primaryClass = {astro-ph.SR},
       adsurl = {https://ui.adsabs.harvard.edu/abs/2019AJ....158...77C}
}

@ARTICLE{Henon1970,
       author = {{Henon}, M.},
        title = "{Numerical exploration of the restricted problem. VI. Hill's case: Non-periodic orbits.}",
      journal = {\aap},
         year = 1970,
        month = nov,
       volume = {9},
        pages = {24-36},
       adsurl = {https://ui.adsabs.harvard.edu/abs/1970A&A.....9...24H}
}

@ARTICLE{Weatherford2026,
       author = {{Weatherford}, Newlin C. and {Bonaca}, Ana},
        title = "{Kinematics of Stellar Streams from Globular Clusters Depend on Black Hole Retention and Star Mass: A Selection Effect for Dark Matter Inference}",
      journal = {\apj},
         year = 2026,
        month = jan,
       volume = {997},
       number = {1},
          eid = {90},
        pages = {90},
          doi = {10.3847/1538-4357/ae21e0},
archivePrefix = {arXiv},
       eprint = {2509.15307},
 primaryClass = {astro-ph.GA},
       adsurl = {https://ui.adsabs.harvard.edu/abs/2026ApJ...997...90W}
}

@ARTICLE{Grondin2024,
       author = {{Grondin}, Steffani M. and {Webb}, Jeremy J. and {Lane}, James M.~M. and {Speagle}, Joshua S. and {Leigh}, Nathan W.~C.},
        title = "{A catalogue of Galactic GEMS: Globular cluster Extra-tidal Mock Stars}",
      journal = {\mnras},
         year = 2024,
        month = mar,
       volume = {528},
       number = {3},
        pages = {5189-5211},
          doi = {10.1093/mnras/stae203},
archivePrefix = {arXiv},
       eprint = {2310.09331},
 primaryClass = {astro-ph.GA},
       adsurl = {https://ui.adsabs.harvard.edu/abs/2024MNRAS.528.5189G}
}

@ARTICLE{Bonaca2025,
       author = {{Bonaca}, Ana and {Price-Whelan}, Adrian M.},
        title = "{Stellar streams in the Gaia era}",
      journal = {\nar},
         year = 2025,
        month = jun,
       volume = {100},
          eid = {101713},
        pages = {101713},
          doi = {10.1016/j.newar.2024.101713},
archivePrefix = {arXiv},
       eprint = {2405.19410},
 primaryClass = {astro-ph.GA},
       adsurl = {https://ui.adsabs.harvard.edu/abs/2025NewAR.10001713B}
}

@ARTICLE{Piatti2020,
       author = {{Piatti}, Andr{\'e}s E. and {Carballo-Bello}, Julio A.},
        title = "{The tidal tails of Milky Way globular clusters}",
      journal = {\aap},
         year = 2020,
        month = may,
       volume = {637},
          eid = {L2},
        pages = {L2},
          doi = {10.1051/0004-6361/202037994},
archivePrefix = {arXiv},
       eprint = {2004.11747},
 primaryClass = {astro-ph.GA},
       adsurl = {https://ui.adsabs.harvard.edu/abs/2020A&A...637L...2P}
}
\bibliographystyle{aasjournalv7}



\end{document}